%% file: cbree_article.tex
\documentclass[hidelinks,onefignum,onetabnum]{siamart220329}

\usepackage{amsfonts} 
\usepackage{enumitem} 
\usepackage{bm} 
\usepackage{dsfont} 
\usepackage{csquotes} 
\usepackage{graphicx}
\usepackage{wrapfig} 
\usepackage{epstopdf}
\usepackage{algorithmic}
\Crefname{ALC@unique}{Line}{Lines} 
\usepackage{hyperref}
\usepackage[shortcuts]{extdash} 
\usepackage{amsopn}
\DeclareMathOperator{\diag}{diag}
\DeclareMathOperator{\MSE}{MSE}
\DeclareMathOperator{\cost}{cost}
\DeclareMathOperator{\relEff}{relEff}

\DeclareMathOperator{\law}{Law}
\DeclareMathOperator{\cov}{Cov}
\DeclareMathOperator{\variation}{C.O.V.}
\DeclareMathOperator{\var}{Var}
\DeclareMathOperator{\supp}{supp}
\DeclareMathOperator{\mattovec}{Vec}
\newcount\colveccount
\newcommand*\colvec[1]{
	\global\colveccount#1
	\begin{bmatrix}
		\colvecnext
	}
	\def\colvecnext#1{
		#1
		\global\advance\colveccount-1
		\ifnum\colveccount>0
		\\
		\expandafter\colvecnext
		\else
	\end{bmatrix}
	\fi
}
\usepackage{tikz}
\usetikzlibrary{shapes,arrows,calc}
\usetikzlibrary{shapes,arrows, positioning}

\newcommand{\vertical}[3]{
	\node (0) at (0,0){};
	\xdef\D{#3}
	\foreach \nodetext[count=\i] in {#1} {
		\xdef\totalIt{\i}
	}
	\foreach \nodetext[count=\i] in {#1} {
		\pgfmathsetmacro\myprev{int{\i-1}}
        \pgfmathsetmacro\curydispl{-1*(\i-1)*\D}
		\pgfmathsetmacro\curxdispl{0}
		\path  ($(\curxdispl cm,\curydispl cm)$) node[block,anchor=center] (\i) {\nodetext};
	}
	
	\foreach \nodetext[count=\i] in {#2}{
		\pgfmathsetmacro\mynext{\ifnum\i<\totalIt int(\i+1)\else 1\fi}
		\ifnum\i<\totalIt \path[draw,-latex] (\i) to  node[fill=white,anchor=center, align=center]{\nodetext} (\mynext) ;\fi
	}
}

\ifpdf
  \DeclareGraphicsExtensions{.eps,.pdf,.png,.jpg}
\else
  \DeclareGraphicsExtensions{.eps}
\fi

\newsiamremark{remark}{Remark}
\newsiamremark{hypothesis}{Hypothesis}
\crefname{hypothesis}{Hypothesis}{Hypotheses}
\newsiamthm{claim}{Claim}

\newcommand{\newindthm}[2]{
    \theoremstyle{plain}
    \theoremheaderfont{\normalfont\sc}
    \theorembodyfont{\normalfont\itshape}
    \theoremseparator{.}
    \theoremsymbol{}
    \newtheorem{#1}{#2}
}
\newindthm{assm}{Assumption}

\headers{Consensus-Based Rare Event Estimation}{K. Althaus, I. Papaioannou, E. Ullmann}

\title{Consensus-Based Rare Event Estimation}
\author{Konstantin Althaus\thanks{Department of Mathematics, Technical University of Munich, Boltzmannstra{\ss}e 3, D-85748 Garching, Germany (\email{konstantin.althaus@tum.de}, \email{elisabeth.ullmann@tum.de}).}
\and Iason Papaioannou\thanks{Engineering Risk Analysis Group, Technical University of Munich, Theresienstra{\ss}e 90, D-80333 Munich, Germany (\email{iason.papaioannou@tum.de}).}
\and Elisabeth Ullmann\footnotemark[1]
}

\ifpdf
\hypersetup{
  pdftitle={Consensus-Based Rare Event Estimation},
  pdfauthor={K. Althaus, I. Papaioannou, E. Ullmann}
}
\fi

\begin{document}

\maketitle

\begin{abstract}
    \input{abstact.tex}									
\end{abstract}

\begin{keywords}
    reliability analysis, importance sampling, McKean--Vlasov stochastic differential equation, Laplace approximation, exponential Runge--Kutta method
\end{keywords}

\begin{MSCcodes}
60H10, 62L12, 65C30, 65N30	
\end{MSCcodes}

\input{introduction.tex}

\input{background.tex}
\input{algorithm.tex}
\input{numerical_experiments.tex}

\input{conclusion.tex}

\bibliographystyle{siamplain}
\bibliography{references}
\end{document}

%% file: abstact.tex
In this paper, we introduce a new algorithm for rare event estimation based on adaptive importance sampling. 
We consider a smoothed version of the optimal importance sampling density, which is approximated by an ensemble of interacting particles.
The particle dynamics is governed by a McKean--Vlasov stochastic differential equation, which was introduced and analyzed in (Carrillo et al., \textit{Stud. Appl. Math.} 148:1069--1140, 2022) for consensus\--based sampling and optimization of posterior distributions arising in the context of Bayesian inverse problems.
We develop automatic updates for the internal parameters of our algorithm.
This includes a novel time step size controller for the exponential Euler method, which discretizes the particle dynamics.
The behavior of all parameter updates depends on easy to interpret accuracy criteria specified by the user.
We show in numerical experiments that our method is competitive to state-of-the-art adaptive importance sampling algorithms for rare event estimation, namely a sequential importance sampling method and the ensemble Kalman filter for rare event estimation.

%% file: introduction.tex
\section{Introduction}
\label{sec:introduction}
In reliability assessments of technical systems, it is often crucial to estimate the probability of failure of the system.
For a system performance depending on $d$-dimensional uncertain inputs, the notion of failure is encoded in a \textit{limit state function} (LSF) $G:\mathbb{R}^d \rightarrow \mathbb{R}$, whose positive values indicate safe system states and non-positive values indicate failure states.
If the uncertain inputs follow some distribution that has the density $\pi$ with respect to the Lebesgue measure on $\mathbb{R}^d$, then the probability of failure $P_f$ is defined as the probability mass of all failure states.
Namely,
\begin{equation}
	\label{eq:def_pf}
	P_f =\int_{G\leq 0}\pi(x)dx.
\end{equation}
\begin{assm}
	\label{rem:initial_distribution}
	In \eqref{eq:def_pf} $\pi$ is the density of the standard normal distribution. 
\end{assm}
This is a common assumption, as in practice one can often find a computable transformation $T$, such that replacing $G$ by $G \circ T$ and $\pi$ by the standard normal density in \eqref{eq:def_pf} does not change the probability of failure, cf. \cite{hohenbichler1981non,Kiureghian1986}.

In safety-critical engineering applications, the probability of failure is small, i.e., failure of the system is a rare event.  Moreover, the evaluation of the limit state function is often computationally expensive, e.g., when the system response is modeled through the numerical solution of a partial differential equation.
Thus the estimation of $P_f$ with crude Monte Carlo is often intractable.
For this reason, several alternative methods for rare event estimation have been developed.
Approximation methods, for example the first-order reliability method (FORM) \cite{HasoferLind:1974,Kiureghian:2004}, determine the most likely failure point and approximate the surface of the failure domain $\{G \leq 0 \}$ near this point by a suitable Taylor expansion.
The probability of failure is then estimated by the probability mass of the approximate failure domain.
Sampling-based methods aim at improving the Monte Carlo estimate of $P_f$ in \eqref{eq:def_pf} through reducing its variance.
Prominent variance reduction methods are importance sampling methods, such as \textit{sequential importance sampling} (SIS) \cite{Papaioannou2016}, and cross entropy--based importance sampling \cite{rubinstein2016simulation,Papaioannou2019}, and splitting methods such as subset simulation \cite{au2001estimation,botev2012efficient}.
Recently, a sampling method termed \textit{Ensemble Kalman Filter for rare event estimation} (EnKF) has been proposed that simulates the dynamics of a stochastic differential equation to obtain failure samples \cite{Wagner2022}.
Here the stochastic dynamics of the ensemble Kalman filter for Bayesian inverse problems (EKI)  \cite{Iglesias2013, Schillings2017} is modified and used to move a sample from the density $\pi$ towards the surface of the failure domain.
The resulting sample then forms the basis for an importance sampling step to estimate $P_f$.

In this paper we propose a new sampling method for rare event estimation.
Our method is similar to the EnKF method in \cite{Wagner2022}, as it also produces a sample for importance sampling by simulating the dynamics of a certain \textit{stochastic differential equation} (SDE).
The SDE has been proposed in \cite{Carrillo2022} for sampling from the posterior distribution in Bayesian inverse problems using a consensus-building mechanism.
The algorithm in \cite{Carrillo2022} is called consensus\--based sampling.
The process of consensus building has been known for a long time \cite{Strogatz2000}, and has recently experienced a new wave of interest in the literature in the context of derivative-free optimization methods \cite{Totzeck2021}.
In particular, advances have been made to make the often heuristic methods amenable to rigorous mathematical proofs on the convergence properties of these methods.
A successful ansatz considers the mean field limit, i.e., the limit of the number of particles $J \rightarrow \infty$ and studies the resulting distribution.
It is then possible to show that a consensus\--based optimization algorithm converges to the global minimizer of a given objective function if certain conditions are fulfilled \cite{Carrillo2018, Fornasier2021}.
Similar theoretical results about the SDE we consider in this paper are proved by the authors of \cite{Carrillo2022}.
We will use their results as the motivation for our algorithm.
 
Our contributions are the following.
We show how consensus\--based sampling can be used for rare event estimation, develop the details of the resulting algorithm, and study its performance in comparison to SIS in \cite{Papaioannou2016} and to the EnKF in \cite{Wagner2022}.
The algorithm has two main ingredients, namely consensus\--based sampling and importance sampling.
The latter is not only used for producing the final estimate of 
$P_f$ in \eqref{eq:def_pf} but also informs the concrete application of consensus\--based sampling.
\subsection{Outline} 
We review importance sampling in Section~\ref{sec:importance_sampling}.
The consensus-based sampling dynamics is discussed in Section \ref{sec:consensus_based_sampling}, including a time-discrete particle approximation of the associated SDE.
In Section \ref{sec:algorithm} we present our algorithm, including the main idea in subsection \ref{sec:underlying_idea}, the parameters of the algorithm in subsection \ref{sec:overview_hyperparameters}, and the automatic parameter tuning in subsections \ref{sec:choosing_smooothing_param}--\ref{sec:stepsize}. 
In Section \ref{sec:enkf_vs_cbree} we compare our algorithm to the EnKF method in \cite{Wagner2022}. 
Finally, in Section \ref{sec:numerical_experiments} we present several numerical experiments with benchmark problems.

\subsection{Notation} \label{sec:notation}
Before we start, let us introduce some notation used throughout the rest of the paper.
Let $i=1:N$ be the shorthand of $i \in \{1,2,\ldots,N-1,N\}$ for some positive integer $N$.
A sample $(x_j)_{j=1:J} \subset \mathbb{R}^d$ is denoted by $\bm{x}$. If the concrete number of of sample points $J \in \mathbb{N}^+$ does not matter we introduce the variables $x_1, \ldots, x_J \subset \mathbb{R}^d$ by their shorthand $\bm{x}\subset \mathbb{R}^d$.
Somewhat inaccurately, we will also talk about the distribution of a sample and write \enquote{$\bm{x} \sim \mu$} for some distribution $\mu$.
What we mean by this formulation is that there are $J$ $\mathbb{R}^d$-valued  random variables $X_1, \ldots, X_J$ defined on some probability space $(\Omega,\mathcal{F},\mathbb{P})$ and a $\omega \in \Omega$ such that $\law(X_j) = \mu$  and $X_j(\omega)=x_j$ for $j=1:J$.
Furthermore, we call a sample i.i.d. if the underlying random variables $X_1, \ldots, X_J$ are statistically independent. 
If the only thing we know and care about a specific distribution is its density, we will denote the distribution by the density itself.
Furthermore, given a symmetric positive definite matrix $A \in \mathbb{R}^{d \times d}$ and a vector $x \in \mathbb{R}^d$ we denote the Euclidean norm, $|\cdot|$, of $A^{-1/2}x$ by $|x|_A$.
We use $A>0$ to denote a positive definite matrix $A$. 
The identity matrix in $d$ dimensions is denoted by $I_d$.
$\mathcal{N}(a,A)$ denotes a Gaussian random vector with mean vector $a$ and covariance matrix $A$.
Finally, the indicator function of a set $S \subset \mathbb{R}^d$ is denoted by $\mathds{1}_S$.

%% file: background.tex
\section{Importance Sampling}
\label{sec:importance_sampling}
Importance sampling (see e.g. \cite[Ch. 9]{Owen2013}) is a modification of Monte Carlo sampling that can yield estimates with a smaller coefficient of variation.
The coefficient of variation, defined in \eqref{eq:def_coeff_of_var} is the natural performance metric in the context of rare event estimation as it can be readily estimated.
It also coincides with the relative $L_2$ error of the estimate if the estimate is unbiased.
Sometimes the relative $L_2$ error is also called the relative root mean square error.

Let us highlight the deficiencies of Monte Carlo sampling and how importance sampling addresses them.
Let $\bm{x} \sim \pi$ be a sample for the density $\pi$.
The Monte Carlo estimate of $P_f$ in \eqref{eq:def_pf} based on $\bm{x}$ is given by
\begin{equation}
	\label{eq:def_mc_estimate}
	\widehat{P}_f^{\textup{MC}} :=  \frac{1}{J} \sum_{j=1:J} \mathds{1}_{\{G \leq 0\} }(x_i).
\end{equation}
If the sample points in $\bm{x}$ are independent, then the coefficient of variation can be computed explicitly:
\begin{equation}
	\label{eq:def_coeff_of_var}
	\variation(\widehat{P}_f^{\textup{MC}}) 
	:= \frac{\sqrt{\var[\widehat{P}_f^{\textup{MC}}]}}{\mathbb{E}[\widehat{P}_f^{\textup{MC}}]} 
	= \frac{\sqrt{\mathbb{E}[(\widehat{P}_f^{\textup{MC}} - P_f)^2]}}{P_f}
	= \sqrt{\frac{1-P_f}{JP_f}}.
\end{equation}
Thus, if $P_f$ is small, say $10^{-k}$, we would need $10^{k+2}$ samples to achieve a coefficient of variation of $10\%$.
Importance sampling aims at reducing the variance of the estimate in \eqref{eq:def_mc_estimate} by sampling from an alternative density $\mu$ instead of the density $\pi$.
The only necessary condition on the new density to ensure that the estimator remains unbiased is  $\supp(\mu) \subseteq \supp(\pi)$.
The new estimate with $\bm{x} \sim \mu$ reads
\begin{equation}
	\label{eq:def_is_estimate}
	\widehat{P}_f^{\textup{IS}} = \frac{1}{J} \sum_{j=1:J}\frac{\pi(x_j)\mathds{1}_{\{G \leq 0\} }(x_i)}{\mu(x_j)} .
\end{equation}
The selling point of importance sampling is the new degree of freedom $\mu$.
We can choose $\mu$ in such a way that  $\variation(\widehat{P}_f^{\textup{IS}})$ becomes small.
There is an optimal choice of $\mu$ leading to a vanishing coefficient of variation.
Namely, using the so-called optimal importance sampling density
\begin{equation}
	\label{eq:def_opt_is_density}
	\mu_{\textup{opt}}(x)	:= \frac{\pi(x)\mathds{1}_{\{G \leq 0\} }(x)}{P_f}
\end{equation}
yields $\var[\widehat{P}_f^{\textup{IS}}] = 0$.
Unfortunately, $\mu_{\textup{opt}}$ is computationally unavailable since we would need to know the value of $P_f$ and the failure domain $\{G\leq 0\}$.
Nonetheless, many algorithms, and indeed the algorithm we present in Section~\ref{sec:algorithm}, sample from a computationally available approximation of $\mu_{\textup{opt}}$.

\section{Consensus--Based Sampling}
\label{sec:consensus_based_sampling}
In the following, we introduce the dynamics of consensus\--based sampling as presented in \cite{Carrillo2022}.
Let $f:\mathbb{R}^d \rightarrow \mathbb{R}$ be a continuous function.
We call $f$ the \emph{energy function}, which is motivated by the  nomenclature used in physics for the exponent of the Gibbs distribution \cite[Ch. 15.9.3]{Ibe2013}.
The consensus\--based sampling dynamics are given by the following McKean--Vlasov stochastic differential equation,
\begin{align}
    dX_t      & = (-X_t + m_\beta(\law(X_t)))dt + \sqrt{2}c_beta(\law(X_t)) dW_t, \label{eq:def_cbs_sde}                 \\
	m_\beta(\nu)   & = \int_{\mathbb{R}^d} \frac{y e^{-\beta f(y)}}{Z} d\nu(y),                 \\
	c_\beta(\nu)^2 & =(1+\beta)\left[ \int_{\mathbb{R}^d} yy^T\frac{e^{-\beta f(y)}}{Z} d\nu(y) 
	- m_\beta(\nu) m_\beta(\nu)^T\right], \label{eq:def_cbs_sde_cov}\\
	Z & = \int_{\mathbb{R}^d} e^{-\beta f(y)} d\nu(y),
\end{align}
where $W_t$ is a $d$-dimensional Wiener process, $m_\beta(\nu)$ is the reweighted mean, and  $c_\beta(\nu)^2$ is the reweighted and scaled covariance of the probability measure $\nu$.
Finally, $\beta >0$ is a parameter called the inverse temperature. 

The authors of \cite{Carrillo2022} show that, given certain conditions on $f$ and $\beta$, the process $X_t$ in \eqref{eq:def_cbs_sde} has an equilibrium distribution which is arbitrarily close to the Laplace approximation associated with the density $\tau \propto e^{-f}$. 
We want to provide an intuition for this result.
The integral in  $m_\beta(\nu)$ weights each $y \in \supp(\nu)$ according to $\nu$ itself times a factor proportional to $e^{-\beta f(y)}$.
Thus, as $\beta$ increases, the probability mass of $X_t$ is concentrated around the minimizer of $f$ on $\supp(\nu)$.
Now we can think of the drift $-X_t + m_\beta(\law(X_T))$ as a vector pointing from the current position of the process towards the minimizer of $f$.
In the limit $\beta \rightarrow \infty$
, this behavior is known as the Laplace principle, \cite[Thm. 4.3.1]{Dembo2010};
for a quantitative formulation of the phenomenon see  \cite[Prop. 21]{Fornasier2021}.
The diffusion $ c_\beta(\law(X_t))$ on the other hand ensures that the covariance of $X_t$ approximates the curvature of $f$ at the minimizer of $f$.
In particular, we note that the reweighting of $\law(X_t)$ in \eqref{eq:def_cbs_sde_cov} requires the scaling factor $1+\beta$ to avoid a degenerate diffusion term for $\beta \rightarrow \infty$, cf. \cite[Prop. 3]{Carrillo2022}.

Let us now make the above intuition more rigorous by stating some of the results from \cite{Carrillo2022}.
Firstly, if $f$ is of the form $1/2 |x-a|_A^2$ (i.e., if $\tau$ is Gaussian) and $\law(X_0)$ is  Gaussian as well, then $X_t$ converges in distribution to $\mathcal{N}(a,A)$ for $t \rightarrow \infty$, cf. \cite[Prop. 3]{Carrillo2022}.
In our application we will deal with a non-quadratic energy function $f$, see Section \ref{sec:underlying_idea}.
For this setting, the authors of \cite{Carrillo2022} provide a convergence result for the case $d=1$, and conjecture that convergence holds for $d>1$ as well.
If the density $\tau$ has the Laplace approximation $\hat{\tau}$,
that is $\hat{\tau}$ is the Gaussian centered at the global minimum $x^*$ of $f$ with covariance $f^{\prime\prime}(x^*)^{-1}$, then the process $X_t$ solving \eqref{eq:def_cbs_sde} converges locally to an equilibrium distribution that is arbitrarily close to $\hat{\tau}$. We restate this result in Theorem~\ref{thm:approx_non_gaussian_target}.

\begin{assm}\label{assm:f}
The energy function $f: \mathbb{R} \rightarrow \mathbb{R}$ satisfies the following conditions:
	\begin{enumerate}[label=(\roman*)]
		\item $f$ is smooth. 
		\item There is a $l > 0$ such that $f^{\prime\prime}(x) > l$ for all $x \in \mathbb{R}$.
		\item For all $i \in \mathbb{N}$ there is a $\lambda_i \in \mathbb{R}$ such that $\sup_{x \in \mathbb{R}} e^{-\lambda_i x^2} |f^{(i)}(x)| < \infty$.
	\end{enumerate}
\end{assm}
\begin{theorem}[Approximation of non--Gaussian Target, {\cite[Thm. 3]{Carrillo2022}}]
	\label{thm:approx_non_gaussian_target}
	Let $f$ satisfy the conditions in Assumption~\ref{assm:f}.
	Then the target distribution $\tau \propto e^{-f}$ has the Laplace approximation $\mathcal{N}(a,A)$ with
	\begin{equation}
		\label{eq:laplace_approx_of_target}
		a = x^*, \quad A  = \frac{1}{f^{\prime\prime}(x^*)}>0.
	\end{equation}
	Let the process $X_t$ be a solution of the SDE $\eqref{eq:def_cbs_sde}$ with the initial distribution $X_0 \sim \mathcal{N}(a_0,A_0)$.
	Furthermore, assume that $A_0 > 0$ and that the initial distribution is close to the Laplace approximation $\mathcal{N}(a,A)$ in \eqref{eq:laplace_approx_of_target}, that is,
		      \begin{equation*}
		      	\left | \colvec{2}{a_0}{A_0} - \colvec{2}{a}{A} \right | \leq r <  A.
		      \end{equation*}
	Then there is a constant $K > 0$ and a constant $\beta_0 > 0 $ depending on $f$ and $r$ such that for any $\beta \geq \beta_0$ the SDE \eqref{eq:def_cbs_sde} has an equilibrium distribution $\mathcal{N}(a_\infty, A_\infty)$ with
		      \begin{equation*}
		      	\left | \colvec{2}{a_\infty}{A_\infty} - \colvec{2}{a}{A} \right | \leq \frac{K}{\beta}.
		      \end{equation*}
		Moreover, $X_t$ converges to the equilibrium distribution if $\beta$ is sufficiently large,
		      \begin{equation}
		      	\label{eq:cbs_sde_conv_to_equlibirum}
		      	\left | \colvec{2}{\mathbb{E}[X_t]}{\cov[X_t]} - \colvec{2}{a_\infty}{A_\infty} \right |
		      	\leq e^{-\left ( 1- \frac{2 K}{\beta}\right )t} \left | \colvec{2}{a_0}{A_0} - \colvec{2}{a_\infty}{A_\infty}  \right |.
		      \end{equation}
	
\end{theorem}

\subsection{Time Discrete Particle Approximation}
\label{sec:particle_approximation}
To take advantage of the theoretical insights we have cited above, we need a computationally available approximation of the dynamics in \eqref{eq:def_cbs_sde}.
In this, we will also follow the authors of \cite{Carrillo2022}. 
They use a time discretization reminiscent of the exponential Euler method for ODEs, cf. \cite{Hochbruck2005}, and the Euler--Maruyama Method for SDEs, cf. \cite[Ch. 9.1]{Kloeden1992}.
Let us call the discretization from \cite{Carrillo2022} the \emph{exponential Euler--Maruyama} method and derive it from the exact solution formula of \eqref{eq:def_cbs_sde}, cf.  \cite[Ch. 4.2]{Kloeden1992}. 
If $X_t$ is the solution of \eqref{eq:def_cbs_sde} on the interval $[t, t+h]$, we have
\begin{align*}
	X_{t+h} & = e^{-h}X_t + \int_{t}^{t+h} e^{s-t-h}m_\beta(\law(X_s)) ds  + \sqrt{2} \int_{t}^{t+h} e^{s-t-h}c_{\beta}(\law(X_s)) dW_s.
\end{align*}
If we evaluate the coefficients $m_\beta$ and $c_\beta$ in the integrals only at time $t$ (like in a normal Euler step), the right hand side reads
\begin{align*}
	{} & e^{-h}X_t + m_\beta(\law(X_t)) \int_{t}^{t+h} e^{s-t-h} ds  + \sqrt{2} c_{\beta}(\law(X_t)) \left ( \int_{t}^{t+h} e^{2(s-t-h)} ds \right )^{1/2} \xi\\
	{} & = e^{-h}X_t +(1-e^{-h}) m_\beta(\law(X_t)) +   \sqrt{1-e^{-2h}} c_{\beta}(\law(X_t)) \xi.
\end{align*}
Here we used the fact that $\int_{t}^{t+h}g(s) dW_s \stackrel{d}{=} \mathcal{N}\left (0,\int_{t}^{t+h}g(s) ^2 ds \right )$ for any square integrable function $g:\mathbb{R} \rightarrow \mathbb{R}$ and introduced the random variable $\xi \sim \mathcal{N}(0,I_d)$.
This yields the stochastic difference  equation studied in \cite{Carrillo2022}:
\begin{equation}
	\begin{split}
		\label{eq:def_disc_cbs_sde}
		X^0 &= X_0, \\
		X^{n+1}	 &= \alpha X^n +(1-\alpha) m_\beta(\law(X^n)) +   \sqrt{1-\alpha^2} c_{\beta}(\law(X^n)) \xi^n,\\
		(\xi^n)_{n \in \mathbb{N}} & \stackrel{\textup{i.i.d.}}{\sim}\mathcal{N}(0,I_d).
	\end{split}
\end{equation}
Here we set $\alpha = e^{-h}$ where $h>0$ is the (fixed) time step size of the discretization.

As the underlying SDE is a McKean--Vlasov SDE, we need to add a second discretization step, which replaces $\law(X^n)$ by the empirical law of a particle ensemble.
Thus, we fix some $J \in \mathbb{N}^+$ and define the following particle approximation of \eqref{eq:def_disc_cbs_sde},
\begin{equation}
	\begin{split}
		\label{eq:cbs_ensemble_update}
		\bm{x}^0  & \stackrel{\textup{i.i.d.}}{\sim} X_0, \\
		x_j^{n+1} & = \alpha x_j^{n} + (1-\alpha)m_\beta(\bm{x}^n) +   \sqrt{1-\alpha^2} c_{\beta}(\bm{x}^n) \xi^n_j, \quad j = 1:J, \\
		(\xi^n_j)_{n \in \mathbb{N} , j = 1:J} & \stackrel{\textup{i.i.d.}}{\sim}\mathcal{N}(0,I_d).
	\end{split}
\end{equation}
Here, with some abuse of notation, we define the coefficients $m_\beta(\bm{x})$ and $c_\beta(\bm{x})$ of the sample as the coefficients of \eqref{eq:def_cbs_sde} evaluated in the empirical measure $1/N \sum_{j=1:J} \delta_{x_j}$, where $\delta_x$ is the Dirac measure in the point $x$.
For the coefficients this means
\begin{equation}
	\label{eq:cbs_ensemble_coefficients}
	\begin{split}
		m_\beta(\bm{x}) &= \sum_{j=1:J}\frac{e^{-\beta f(x_j)}}{\sum_{k=1:J}e^{-\beta f(x_k)}} x_j, \\
		c_{\beta}(\bm{x})^2 &= (1+\beta)\left(\sum_{j=1:J}\frac{e^{-\beta f(x_j)}}{\sum_{k=1:J}e^{-\beta f(x_k)}} x_jx_j^T - m_\beta(\bm{x})m_\beta(\bm{x})^T\right).
	\end{split}	
\end{equation}
Importantly, this sequence of ensembles can be computed if we can sample from the initial distribution $\law(X_0)$.
\begin{remark}[Other Theoretical Properties]
	\label{rem:other_theoretical_properties}
	The paper \cite{Carrillo2022}, where the SDE  \eqref{eq:def_cbs_sde} has debuted, does not prove the existence or uniqueness of its solution(s).
	The lack of theoretical results extends also to the approximations \eqref{eq:def_disc_cbs_sde} and \eqref{eq:cbs_ensemble_update}: We do not know if they converge to the original SDE.
	Therefore the paper \cite{Carrillo2022} also proves the following result for the time discrete process \eqref{eq:def_disc_cbs_sde}, cf. \cite[Thm. 3]{Carrillo2022}.
	If the assumptions of Theorem \ref{thm:approx_non_gaussian_target} hold true, 
	we also have for all $n \in \mathbb{N}$:
	\begin{equation}
		\label{eq:disc_cbs_sde_conv_to_equlibirum}
		\left | \colvec{2}{\mathbb{E}[X^n]}{\cov[X^n]} - \colvec{2}{a_\infty}{A_\infty} \right |
		\leq \left ( \alpha + (1-\alpha^2)\frac{K}{\beta} \right)^n
		\left | \colvec{2}{a_0}{A_0} - \colvec{2}{a_\infty}{A_\infty}  \right |.
	\end{equation}
\end{remark}

%% file: algorithm.tex
\section{CBREE Algorithm}
\label{sec:algorithm}
In this section we present the novel algorithm called Consensus\--Based Rare Event Estimation (CBREE).
First, we explain the main idea in Section \ref{sec:underlying_idea}.
Next, we present a pseudocode formulation of CBREE in Algorithm \ref{alg:cbree}.
Finally, we work out the details in Sections \ref{sec:overview_hyperparameters}--\ref{sec:stopping_criterion}.
\subsection{Main Idea}
\label{sec:underlying_idea}

In this section, we explain how to employ consensus\--based sampling for importance sampling.
We assume that the particle approximation \eqref{eq:cbs_ensemble_update} inherits the theoretical properties of the consensus\--based sampling dynamics in Theorem \ref{thm:approx_non_gaussian_target}.
To produce a point estimate of $P_f$ with a small coefficient of variation, we would like to obtain a sample that approximates the optimal importance sampling density $\mu_{\textup{opt}}$ from \eqref{eq:def_opt_is_density}.
For this reason, we choose an appropriate target density $\tau \propto e^{-f}$ in Theorem \ref{thm:approx_non_gaussian_target}.
Assume we have a smooth approximation $I$ of the indicator function $\mathds{1}_{\mathbb{R}^-_0}$ of the form
$	I(x, s) \rightarrow \mathds{1}_{\mathbb{R}^-_0}(x) $ for $s \rightarrow \infty$ \textup{a.e.,}
where $s$ denotes the so-called smoothing parameter.
Then we define our target density as $\tau = \pi(x,s)\propto I(G(x), s) \pi(x)$.
Many choices are conceivable and have been used for the approximation $I$.
While \cite{Papaioannou2016} uses $\Phi(-s x)$ where $\Phi$ is the cumulative distribution function of the one-dimensional standard normal distribution, we have obtained the best results using a transformed logistic function,
\begin{equation}\label{I}
	I(x,s)  = \frac{1}{2}\left(1 -\frac{s x}{\sqrt{s^2x^2+1}} \right).
\end{equation}

\begin{wrapfigure}{R}{.5\textwidth}
    \begin{centering}
            \begin{tikzpicture}
                \tikzstyle{block} = [rectangle, draw, text width=17em, text centered, rounded corners, minimum height=3em, align=center]
                \vertical{
                    {$\mu_{\textup{opt}}$},
                    {$\pi(\cdot, s) \rightarrow \mu_{\textup{opt}}, s \rightarrow \infty$},
                    Laplace approximation {$\mathcal{N}(a,A)$} of {$\pi(\cdot,s)$},
                    Equilibrium distribution {$\mathcal{N}(a_\infty,A_\infty)$ of $X_t$},
                    Time discretization $X^N$ of $X_{Nh}$,
                    Particle approximation {$(x_j^N)_{j=1:J}$ of $\law(X^N)$}
                }{
                    Smoothing parameter {$s$},
                    Assume the existence,
                    {Inverse temperature $\beta$},
                    {Time step size $h$ and number of steps $N$},
                    {Sample size $J$}
                }{2.7}
            \end{tikzpicture}
        \caption{Approximation steps performed by the proposed CBREE method to represent the optimal importance sampling density $\mu_{\textup{opt}}$.}
        \label{fig:approximation_steps}
    \end{centering}
\end{wrapfigure}
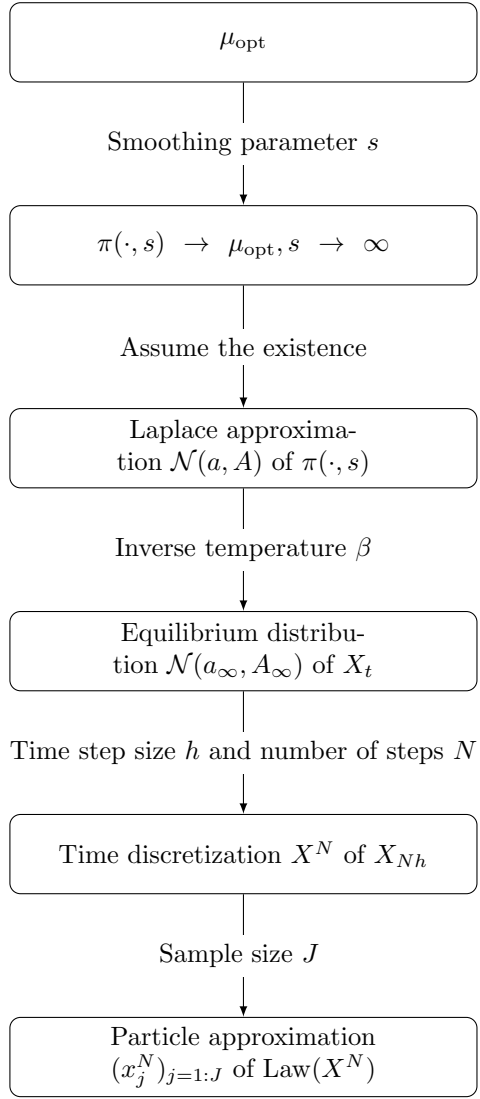

Now we use the consensus\--based sampling dynamics in \eqref{eq:def_cbs_sde} to approximate $\tau$.
That means we employ the energy function $f(x) = -\log I(G(x), s) \pi(x) $ in the particle update of \eqref{eq:cbs_ensemble_update} to transform the initial sample $\bm{x}^0$ in $N$ steps into the ensemble $\bm{x}^N$.
The approximation steps that lead from the density $\mu_{\textup{opt}}$ to the final sample $\bm{x}^N$ are visualized in Figure \ref{fig:approximation_steps}.
As the diagram indicates, the different steps introduce several parameters.
We have  mentioned the smoothing parameter $s >0 $.
The process $X_t$ behind consensus\--based sampling depends on another parameter, the inverse temperature $\beta>0$, cf. Section \ref{sec:consensus_based_sampling}. 
Next, the discretization of the continuous dynamics in time introduces a time step size $h>0$ and number of time steps $N \in \mathbb{N}^+$, giving the time discrete process $(X^n)_{n=1:N}$.
Finally, discretizing $(X^n)_{n=1:N}$ in law entails replacing $X^n$ by an ensemble of $J$ particles $(x_j^n)_{j=1:J} \subset \mathbb{R}^d$, cf. Section \ref{sec:particle_approximation}.
In Sections \ref{sec:overview_hyperparameters}--\ref{sec:stepsize} we will show how these parameters can be chosen adaptively based on a prescribed accuracy of the final estimate of $P_f$ in \eqref{eq:def_pf}.

Now we explain how we can use the final sample $\bm{x}^N$ for importance sampling, cf. \eqref{eq:def_is_estimate}.
For this, we need to know the density $\mu^N$ of the sample. 
If we assume that for a large sample size $J$ we have $\bm{x}^N \sim \mu^N \approx \law(X^N)$, where $X^N$ is the time discrete process in \eqref{eq:def_disc_cbs_sde}, we can use the density of $\law(X^N)$ for importance sampling.
As we furthermore know that $X^n$ is Gaussian for $n>0$ if $X^0$ is Gaussian, cf. \cite[Lem. 2]{Carrillo2022}, we assume that $\mu^N$ is Gaussian and estimate its parameters, namely the empirical mean and covariance of $\bm{x}^N$, to evaluate $\mu^N$. 

Let us make clear which assumptions justify the use of $\bm{x}^N$ with a fitted Gaussian for the importance sampling estimate \eqref{eq:def_is_estimate}:
\begin{itemize}
    \item[(A1)] The smoothed optimal importance sampling density $\pi(x,s)\propto I(G(x), s) \pi(x)$ has a Laplace approximation $\mathcal{N}(a,A)$ for some $s>0$, which is well suited for importance sampling.
    \item[(A2)] The equilibrium distribution  $\mathcal{N}(a_\infty,A_\infty)$ of the SDE \eqref{eq:def_cbs_sde} is close to the Laplace approximation $\mathcal{N}(a,A)$.
   \item[(A3)] The time-discrete particle approximation \eqref{eq:cbs_ensemble_update} has approximately the same equilibrium distribution as the associated SDE in \eqref{eq:def_cbs_sde}.
\end{itemize}
Unfortunately, it is difficult to check (A1) in practice.
We note that (A1) imposes restrictions on the shape of the failure domain $\{G\leq 0 \}$.
If the problem is multimodal, i.e., if $\mu_{\textup{opt}}$ and $\pi(x,s)$ have multiple global maxima, then the Laplace approximation is not well defined.
Hence it is not clear if the particle approximation has an equilibrium distribution and whether this distribution is a good choice for importance sampling.

\subsection{Overview of Parameters}
\label{sec:overview_hyperparameters}
The approach in Section \ref{sec:underlying_idea} introduces several parameters that we need to tune. Our goal is to propose adaptive schemes for each parameter.
Furthermore, we present a stopping criterion that stops the iteration \eqref{eq:cbs_ensemble_update} after $N$ steps.
If we use the proposed energy function $f(x) = -\log I(G(x), s) \pi(x) $, we can think of the particle update \eqref{eq:cbs_ensemble_update} as the function 
\begin{equation}
	\label{eq:cbs_step_fun}
	\texttt{cbs\_step}: \left( \bm{x}^n, s^{n+1}, \beta^{n+1}, h^{n+1}\right) \mapsto \bm{x}^{n+1}.
\end{equation}
In our algorithm the parameter triplet $(s^{n+1}, \beta^{n+1}, h^{n+1})$ will be adjusted before each iteration.
We have settled on the following ideas.
\begin{itemize}
	\item[(P1)] The smoothing parameter $s$ controls the distance between the target distribution $\pi(\cdot, s)$ and the optimal density $\mu_{\textup{opt}}$. We use an adaptive scheme that has been successfully used for other importance sampling methods, e.g. \cite{Papaioannou2016, Wagner2022}.
	\item[(P2)] The inverse temperature $\beta$  shapes the density used to compute the coefficients of the SDE \eqref{eq:def_cbs_sde}. We use an information measurement proposed by \cite{Carrillo2022} to adjust $\beta$.
	\item[(P3)] The stepsize $h$ in \eqref{eq:def_disc_cbs_sde} can be adjusted using the established ideas from adaptive time integrators. We use an \emph{ordinary differential equation} (ODE) as a proxy to construct a custom stepsize controller.
	\item[(P4)] We use an established stopping criterion from \cite{Papaioannou2016, Wagner2022}. Additionally, we present a second stopping criterion that can improve the performance of CBREE.
\end{itemize}
A pseudocode version of the resulting method is given in Algorithm \ref{alg:cbree}.
Let us elaborate on the four items above in the following sections.

\begin{algorithm}
    \caption{Consensus-based sampling for rare event estimation (CBREE)}
    \label{alg:cbree}
    \begin{algorithmic}[1]
        \REQUIRE{\quad \\
            $\bm{x}^0= (x_j^0)_{j=1:J} \subset \mathbb{R}^d$  (initial Gaussian ensemble)\\
            $\epsilon_{\textup{Target}}>0$ (tolerance for stepsize selection)\\
            $\Delta_{\textup{Target}}>0$ (tolerance for convergence check) \\
            $N_{\textup{obs}}\geq 2$ (length of observation window for divergence check)
        }
        \\\hrulefill
        \STATE{Compute initial stepsize $h^1$ according to \eqref{eq:starting_stepsize}}
        \FOR{$n \in \mathbb{N}$}
        \STATE{Compute $\hat{P}_f^n$ from \eqref{eq:cbree_is_esimate} based on $\bm{x}^n$}
        \IF{convergence check is passed, cf. Section \ref{sec:stopping_criterion},}
        \label{alg_line:begin_cbree_vmfn_recycle}
        \RETURN{ $\hat{P}_f^n$}
        \ENDIF
        \IF{$n \geq N_{\textup{obs}}$ and divergence check is passed, cf. Section \ref{sec:stopping_criterion},}
        \label{alg_line:div_check}
        \RETURN{$\frac{1}{N_{\textup{obs}}} \sum_{k=n-N_{\textup{obs}}+1:n}  \hat{P}_f^k$}
        \ENDIF
        \STATE{Compute $s^{n+1}$, cf. Section \ref{sec:choosing_smooothing_param}}
        \STATE{Compute $\beta^{n+1}$, cf. Section \ref{sec:choosing_temperature}}
        \IF{$n \geq 0$ and $n$ is even}
        \STATE{Compute $h^{n+1}$, cf. Section \ref{sec:stepsize}}
        \ELSE
        \STATE{$h^{n+1} \gets h^n$}
        \ENDIF
        \STATE{$\bm{x}^{n+1} \gets \texttt{cbs\_step} \left( \bm{x}^n, s^{n+1}, \beta^{n+1}, h^{n+1}\right)$, cf. \eqref{eq:cbs_step_fun}}	 

        \label{alg_line:end_cbree_vmfn_recycle}
        \ENDFOR
    \end{algorithmic}
\end{algorithm}

\subsection{Choosing the Smoothing Parameter}
\label{sec:choosing_smooothing_param}
Assumption \ref{rem:initial_distribution} can be read as $\bm{x}^0 \sim \pi(\cdot,0) = \pi$.
In Section \ref{sec:underlying_idea} we have also sketched our reasoning why after a  sufficient number of iterations, say $N>0$, the sample $\bm{x}^N$ is approximately distributed according to the Laplace approximation of $\pi(\cdot,s^N) \approx \mu_{\textup{opt}}$.
This suggests increasing $s$ in between the iterations.
On the one hand, we want to increase the parameter $s$ fast enough such that the target distribution $\pi(\cdot,s)$ is close to $\mu_{\textup{opt}}$ after as few iterations as possible.
On the other hand, Theorem \ref{thm:approx_non_gaussian_target} provides only local convergence.
Thus, the difference $s^{n+1}-s^n$ must be small to ensure that the ensemble $\bm{x}^n$  moves towards the correct attractor, the Laplace approximation of $\pi(\cdot,s^{n+1})$.

Our problem of increasing  $s$ adaptively is very similar to a situation that arises in the implementation of sequential importance sampling, cf. \cite{Papaioannou2016}.
There a sequence of samples $\bm{x}^0, \bm{x}^1, \ldots$ with $\bm{x}^n \sim \pi(\cdot,s^n)$ and $s^n < s^{n+1} $ for all $n \geq 0$ is produced.
There as well, a smaller increment $s^{n+1}-s^n$ is associated with more costs and higher accuracy.
Given a user specified tolerance $\Delta_{\textup{Target}}>0$ the authors of \cite{Papaioannou2016} propose to choose $s^{n+1}$ as the  minimizer of the function
\begin{equation}
	\label{eq:opt_prob_cbs_stepsize}
	s \mapsto \left(\widehat{\variation}(\bm{q})- \Delta_{\textup{Target}}\right)^2, \quad q_j = \frac{\pi(x^n_j, s)}{\pi(x^n_j, s^{n})}, \quad j =1:J
\end{equation}
on the domain $[s^n,\infty)$.
	Here $\widehat{\variation}$ is the empirical counterpart to the coefficient of variation defined in \eqref{eq:def_coeff_of_var}.
	This adaptive scheme works, because $\widehat{\variation}(\bm{q})^2$ approximates the  quantity
	\begin{equation}
		\label{eq:delta_distance}
		\Delta(\pi(\cdot, s), \pi(\cdot, s^{n}))^2 := \int_{\mathbb{R}^d}\frac{\pi(x, s)}{\pi(x,s^n)}\pi(x, s)dx - 1.
	\end{equation}
	For two densities ${\mu}$ and ${\nu}$ with $\supp({\nu}) \subseteq \supp({\mu})$, the quantity $\Delta({\mu}, {\nu})$ can be interpreted as a distance between $\mu$ and $\nu$ since the following  relationship with the Kullback--Leibler divergence $D_{\textup{KL}}(\mu,\nu) = \int_{\mathbb{R}^d} \log\left(\frac{\mu(x)}{\nu(x)}\right)\mu(x)dx$ holds:
	\begin{equation*}
		\begin{split}
			\Delta(\mu, \nu)^2 & \geq e^{D_{\textup{KL}}(\mu,\nu)} -1, \\
			\Delta(\mu, \mu)^2 & = e^{D_{\textup{KL}}(\mu,\mu)} -1 = 0.
		\end{split}
	\end{equation*}
	Now we come back to our algorithm.
	If we make the simplifying assumption $\bm{x}^n \stackrel{\textup{approx}}{\sim}\pi(\cdot,s^n) \propto I(G,s^{n})\pi$, i.e., we ignore the fact that $\bm{x}^n$ only approximates the Laplace approximation of $\pi(\cdot,s^n)$, 
	we are in the setting of sequential importance sampling.
	Therefore we propose to also use the update scheme \eqref{eq:opt_prob_cbs_stepsize}.
	We justify ignoring the difference between $\mu^n$ and $\pi(\cdot,s^n)$, as in our experience using the actual densities of $\bm{x}^{n}$ and $\bm{x}^{n+1}$, i.e.,  changing the weights $\bm{q}$ in \eqref{eq:opt_prob_cbs_stepsize} to
	\begin{equation*}
		q_j = \frac{\mu^{n+1}(x^n_j)}{\mu^{n}(x^n_j)},
	\end{equation*}
	does not make our method more accurate but only increases the cost (measured in the number of LSF evaluations).
	The reason for the cost increase is that the evaluation of the objective function from \eqref{eq:def_opt_is_density} in $s$ involves fitting the Gaussian $\mu^{n+1}$ to the result of $\texttt{cbs\_step}\left( \bm{x}^n, s, \beta^{n+1}, h^{n+1}\right)$.
	Finally, our experience also suggests that it is prudent to limit the increase of $s$ in dependence on the stepsize $h > 0$.
	We will explain the reasoning behind this in Section \ref{sec:stepsize}.
	Here we already mention that we introduce a safety parameter $\textup{Lip}(s)>0$ and restrict the domain of the objective function in \eqref{eq:opt_prob_cbs_stepsize}  to $[s^n, s^n+\textup{Lip}(s)h]$.
	We will mostly use the choice $\textup{Lip}(s)=1$.
	\subsection{Choosing the Inverse Temperature}
	\label{sec:choosing_temperature}
	From Theorem \ref{thm:approx_non_gaussian_target} we know that the inverse temperature $\beta$ should be larger than some $\beta_0$ whose value is unknown in practice.
	Instead we follow an update strategy for $\beta$ developed 	by authors of the original consensus\--based sampling algorithm, \cite{Carrillo2022}.
	We should think of the ensemble $\bm{x}^n$ in \eqref{eq:cbs_ensemble_update} in combination with the weights $w_j^n= \left [I(G(x_j^n),s^{n+1})\pi(x_j^n)\right ]^{\beta^{n+1}}$ from \eqref{eq:cbs_ensemble_coefficients} as a weighted sample.
	In this context, a larger $\beta^{n+1}$ assigns to the point with the biggest value of $I(G(x),s^{n+1})\pi(x)$ a higher weight with respect to the remaining points.
	If the distribution of the weights $\bm{w}^n$ is too skewed, only a few LSF evaluations
	are effectively contributing to the update formula of the ensemble.
	This can have adverse effects on the convergence of the consensus\--based sampling algorithm
	because we are not using the full information of all $J$ particles to approximate the current law of the process in \eqref{eq:def_disc_cbs_sde}.
	For this reason, the authors of \cite{Carrillo2022} propose to choose $\beta$ in a way that fixes the effective sample size of the weighted sample.
	The effective sample size of the ensemble  $\bm{x}^n$ weighted by $\bm{w}^n$ is defined as
	\begin{equation}
		\label{eq:def_effective_sample_size}
		J^n_{\textup{ESS}}(\beta) = \frac{\left ( \sum_{j=1:J} \left [I(G(x_j^n),s^{n+1})\pi(x_j^n)\right )^{\beta} \right )^2}{\sum_{j=1:J}\left [I(G(x_j^n),s^{n+1})\pi(x_j^n)\right ]^{2\beta} }.
		\end{equation}
		The authors of \cite{Carrillo2022} show that for each $J^* \in (1, J)$ there is a unique $\beta^* >0$ such that $J_{\textup{ESS}}(\beta^*) = J^*$.
		Having determined the value of $s^{n+1} $ we follow \cite{Carrillo2022} and define $\beta^{n+1}$  as the solution of
		\begin{equation}
			\label{eq:beta_update}
			J^n_{\textup{ESS}}(\beta)  = \frac{J}{2}.
		\end{equation}
		\subsection{Choosing the Stepsize}
		\label{sec:stepsize}
		The stepsize $h>0$ yields the approximation $(X^n)_{n\in \mathbb{N}}$ defined in \eqref{eq:def_disc_cbs_sde} of the continuous process $(X_t)_{t\geq 0}$ from \eqref{eq:def_cbs_sde}.
		In the theory of discretizing ODEs, the subject of stepsize control is fairly advanced, see e.g. \cite[Ch. II.4]{ErnstHairer1993}.
		Stepsize selection in the context of SDEs, on the other hand, is more involved as one needs to sample a stochastic process:
		If a timestep is rejected, the ensuing resampling of the process has to take into account the earlier samples, cf.  \cite[Sec. 4]{Burrage2003}.
		Furthermore, we would like to work with the original time discretization scheme \eqref{eq:def_disc_cbs_sde} to make use of the convergence result \eqref{eq:disc_cbs_sde_conv_to_equlibirum}.
		To our knowledge, there is no adaptive exponential Euler method for SDEs. 
		For these reasons, we have developed an unconventional stepsize control for the application at hand. 
		
		Firstly, we show that there is a readily available ODE discretization whose stepsize we can control instead of directly controlling the stepsize of the SDE discretization.
		Secondly, we construct a higher order method operating on the grid induced by the auxiliary stepsize $\widehat{h} := 2h$.
		At this point, we can use the standard approach to estimate the optimal stepsize \emph{for every second timestep}.
		Thirdly, we explain in what aspects our custom stepsize controller deviates from the standard approach.
		Finally, we also tackle the problem of finding a suitable initial stepsize.
		The remainder of the section is organized into paragraphs that deal with the steps above one by one.
		\paragraph{The Proxy Discretization}
		We would like to use the ODE theory of stepsize control.
		Thus, we need an ODE discretization that can be used as a proxy for controlling the parameter $h$ in the SDE discretization.
		We consider the dynamics of the first two moments of $X_t$ and $X^n$, from \eqref{eq:def_cbs_sde} and \eqref{eq:def_disc_cbs_sde} respectively, for this purpose.
		If we take the initial value prescribed by Assumption~\ref{rem:initial_distribution}, i.e., $\law(X_0) = \law(X^0) = \mathcal{N}(0,I_d)$,  into account, we know that $X_t$ and $X^n$ are Gaussian for $t,n>0$, cf. \cite[Lem. 2]{Carrillo2022}.
		Then we can apply It\^{o}'s lemma, cf. \cite[Thm. 7.4.3]{Kuo2006}, to the processes $X_t$ and $X^n$ and obtain two closed systems.
		The continuous process yields
		\begin{equation}
			\label{eq:ode_moments_of_sde}
			\begin{split}
				\law(X_0) &=\mathcal{N}(0,I_d), \\
				\frac{d\mathbb{E}[X_t]}{dt} &= -\mathbb{E}[X_t] +  m_\beta(\mathcal{N}(\mathbb{E}[X_t],\cov[X_t])), \\
				\frac{d\cov[X_t]}{dt} &= -2\cov[X_t] + 2 c_\beta(\mathcal{N}(\mathbb{E}[X_t],\cov[X_t]))^2.
			\end{split}
		\end{equation}
		The time-discrete process on the other hand gives
		\begin{equation}
			\label{eq:ode_moments_of_disc_sde}
			\begin{split}
				\law(X^0) &=\mathcal{N}(0,I_d), \\
				\mathbb{E}[X^{n+1}] &= \alpha\mathbb{E}[X^n] + (1-\alpha) m_\beta(\mathcal{N}(\mathbb{E}[X^n],\cov[X^n])), \\
				\cov[X^{n+1}] &= \alpha^2\cov[X^n] + (1-\alpha^2) c_\beta(\mathcal{N}(\mathbb{E}[X^n],\cov[X^n]))^2.
			\end{split}
		\end{equation} 
		Now, the crucial point is that one also obtains the recursion \eqref{eq:ode_moments_of_disc_sde} if one approximates the ODE \eqref{eq:ode_moments_of_sde} with the exponential Euler method for ODEs, cf. \cite{Hochbruck2005}.
		The generic form for approximating a semilinear ODE in $m$ dimensions,
		\begin{equation}
			\label{eq:semilinear_ode}
			\dot{x} +Ax=  g(t,x), \quad x(0) = x_0,
		\end{equation}
		with the exponential Euler method reads
		\begin{equation}
		\label{eq:exp_euler}
			x^{n+1} = e^{-hA}x^n + h\phi(-hA)g(t^n,x^n),
		\end{equation}
		where $\phi(-hA) = \int_0^1 e^{-(1-y)hA}dy = \left( e^{-hA} - I_m\right) (-hA)^{-1}$.
		For this reason we consider the sequence 
		\begin{equation*}
			(\theta^n)_{n \in \mathbb{N}} := (\mathbb{E}[X^n], \mattovec(\cov[X^n]))_{n \in \mathbb{N}} \subset \mathbb{R}^m, \text{ with } m = d + d^2,
		\end{equation*}
		that is obtained by first discretizing the SDE \eqref{eq:def_cbs_sde} in time and then taking the first two moments of each iterate as the output of the exponential Euler method applied to the ODE \eqref{eq:ode_moments_of_sde}.
		\paragraph*{Classical Stepsize Control}
		Now we show how to construct a method of higher order using the sequence $(\theta^n)_{n \in \mathbb{N}} $ and develop a classical stepsize controller along the lines of \cite[Ch. II.4]{ErnstHairer1993}.
		Note that in practice the moments of $X^n$ collected in $\theta^n$ are computationally not available.
		Instead, we have only access to the particle approximation \eqref{eq:cbs_ensemble_update} of $X^n$.
		Thus we will replace the moments of $X^n$ with their empirical counterparts in the algorithm.

		Let us state a general form of an explicit exponential Runge--Kutta method with $\ell$ stages for approximating the solution of \eqref{eq:semilinear_ode}:
		\begin{equation}
			\label{eq:general_exponential_rk_method}
			\begin{split}
				x^{n+1}& = e^{-hA}x^n + h \sum_{i=1:\ell}b_i(-hA)G^n_i, \\
				G^n_i &=g\left (t^n + c_ih,e^{-c_ihA}x^n+h\sum_{k=1:i-1}a_{i,k}(-hA)G^n_k\right ).
			\end{split}
		\end{equation}
		Here $ b_i(-hA)$ and $a_{ij}(-hA)$ with $1 \leq i, j \leq \ell$ are matrices depending on the matrix $-hA$ whereas $c \in \mathbb{R}^\ell$. 
		The parameters also come with two consistency conditions \cite[(2.23)]{Hochbruck2005}, namely
		\begin{equation}
			\label{eq:consistency_exp_integrator}
			\sum_{j=1:\ell}b_j(-hA) = \phi(-hA), \quad \sum_{j=1:i-1}a_{ij}(-hA) = c_i \phi(-hc_iA),\quad i=1:\ell.
		\end{equation}
		
		If we consider only every second iteration of $(\theta^n)_{n \in \mathbb{N}}$, we obtain the output of an auxiliary exponential Runge--Kutta method with two stages and stepsize $\widehat{h}=2h$.
		Using a Butcher tableau notation introduced in \cite[Sec. 2.3]{Hochbruck2005}, we write down the parameters of this method in the left tableau of Table \ref{tab:exp_euler_table}.
		We will call the auxiliary method the exponential two-step Euler.
		In the following, we determine its order and construct a method of higher order operating also on the grid $\{2nh; n \in \mathbb{N}\}$, which is the key for stepsize control.

		\begin{table}[h]
			\begin{minipage}{.49\linewidth}
				\centering
				\begin{tabular}{c|cc}
					$0$           & 0                                                               & 0                                              \\ 
                    $\frac{1}{2}$ & $\frac{1}{2}\phi\left(\frac{-hA}{2}\right)$                  & 0                                              \\ 
					\hline 
                                  & $\frac{1}{2}e^{\frac{-hA}{2}}\phi\left ( \frac{-hA}{2}\right )$ & $ \frac{1}{2}\phi\left ( \frac{-hA}{2}\right )$ \\ 
				\end{tabular} 	
			\end{minipage}
			\begin{minipage}{.49\linewidth}
				\centering
			\begin{tabular}{c|cc}
				$0$           & 0                                 & 0                \\ 
                    $\frac{1}{2}$ & $\frac{1}{2}\phi\left(\frac{-hA}{2}\right)$                  & 0                                              \\ 
				\hline 
				              & $\widehat{b}_1(-hA)$                  & $\widehat{b}_2(-hA)$ \\ 
			\end{tabular}
			\end{minipage}
			\caption{Tableau of the auxiliary method obtained by saving only every second exponential Euler step (exponential two-step Euler) and the exponential midpoint rule.}
			\label{tab:exp_euler_table}
		\end{table}
		The exponential two-step Euler method is consistent as it satisfies \eqref{eq:consistency_exp_integrator} and thus is also at least of order 1, cf. Table 2.2 in \cite{Hochbruck2005}.
		But it is also not of a higher order as the second order conditions are not met.
		These conditions are:
		\begin{equation}
			\label{eq:exp_integrators_second_order_condition}
			\begin{split}
            \sum_{i=1:\ell}c_i b_i(-hA) & = e^{-hA}(hA)^{-2} + (hA)^{-1} - (hA)^{-2}, \\
				a_{21}(-hA)& =\frac{1}{2}\phi\left ( \frac{-hA}{2}\right ).
			\end{split}
		\end{equation}
		As we can see from Table \ref{tab:exp_euler_table} the exponential two-step Euler produces $\sum_{i=1:\ell}c_i b_i(-hA) =\frac{1}{4}\phi\left ( \frac{-hA}{2}\right )$.
		Now it is also easy to modify the exponential two-step Euler method without changing the stages $G^n_1$ and $G^n_2$ in \eqref{eq:general_exponential_rk_method} such that the resulting method is of order $2$.
		We replace the functions $b_1(-hA) = \frac{1}{2}e^{\frac{-hA}{2}}\phi\left ( \frac{-hA}{2}\right )$ and $b_2(-hA) = \frac{1}{2}\phi\left ( \frac{-hA}{2}\right )$ in Table \ref{tab:exp_euler_table} by 
		\begin{equation}
			\begin{split}
                \widehat{b}_1(-hA) &:= \phi(-hA) - 2\left(e^{-hA}(hA)^{-2} + (hA)^{-1} - (hA)^{-2}\right), \\
				\widehat{b}_2(-hA) &:=2\left(e^{-hA}(hA)^{-2} + (hA)^{-1} - (hA)^{-2}\right).
			\end{split}
		\end{equation}
		We call the resulting method the exponential midpoint rule as taking the limit $A \rightarrow 0$ recovers the classical midpoint rule.
		From this it is also immediate that this method whose tableau is given in the right tableau of Table \ref{tab:exp_euler_table} is of no higher order than $2$ which is the order of the classical midpoint rule, cf. \cite[Ch. II. 1]{ErnstHairer1993}.
		Now, given the original exponential Euler approximation $(\theta^n)_{n \in \mathbb{N}}$ of the solution of \eqref{eq:ode_moments_of_sde} that is defined on the grid $\{hn; n \in \mathbb{N}\}$ we can compute two different discretizations without any additional evaluations of the right hand side of \eqref{eq:ode_moments_of_sde}, namely 
		\begin{equation}
			\begin{split}
				\psi^n &:= \theta^{2n}, \\
                \phi^n &:= \widehat{h} \cdot \widehat{b}_1(-\widehat{h}A) \theta^{2n-1} + \widehat{h} \cdot  \widehat{b}_2(-\widehat{h}A)\theta^{2n}.
			\end{split}
		\end{equation}
		Note that $(\psi^n)_{n \in \mathbb{N}}$ is the result of the exponential Euler two-step method while $(\phi^n)_{n \in \mathbb{N}}$ is the output of the exponential midpoint rule applied to the ODE \eqref{eq:ode_moments_of_sde} on the grid $\{2hn; n \in \mathbb{N}\}$.
		As the second method is of order two while the former is of order one, we can estimate the local truncation error at every second gridpoint of the grid $\{hn; n \in \mathbb{N}\}$, cf. \cite[Chp. II. 3 \& 4]{ErnstHairer1993}.
		This leads to the well established optimal stepsize estimate \cite[(4.12)]{ErnstHairer1993}, which is based on the user specified absolute and relative tolerance for each solution component $i$: $\epsilon^{\textup{abs}}_i,\epsilon^{\textup{rel}}_i \geq 0$.
		In our case the formula \cite[(4.12)]{ErnstHairer1993} reads
		\begin{equation}
			\label{eq:optimal_stepsize}
			h_{\textup{opt}}^n := \left ( \frac{1}{\textup{err}^n}\right )^{1/2}h^n, \quad \textup{err}^n:= |\phi^n -\psi^n|_\Gamma,
		\end{equation}
		where
		\begin{equation}
			\label{eq:error_norm_weights}
			\Gamma_{i,j} = 
			\begin{cases}
				m\left( \epsilon^{\textup{abs}}_i + \epsilon^{\textup{rel}}_i \max(|\psi^n_i|,|\psi^{n-1}_i|)\right), & i = j,         \\
				0,                                                                                                    & {i\neq j}, \\
			\end{cases}
            \text{ for } i,j=1:m.
		\end{equation}
        As above, we have $m=d^2 + d$.
		This means that after the computation of the $n$th step, one can check if the local error is too big and what would have been the optimal stepsize.
		If the error is too large ($\textup{err}^n>1$ ) one usually recomputes the $n$th step using the optimal stepsize $h_{\textup{opt}}^n$.
		Otherwise one continues with the next step and uses $h_{\textup{opt}}^n$ as the first guess for the new stepsize $h^{n+1}$.
		
        \paragraph*{Custom Stepsize Control}
        This is the point we diverge from the classical theory of stepsize control, which we have followed so far. 
        In our algorithm, we do not recompute steps that one would normally reject.
        The reason for this is that the size of the errors $|\theta^n - (\mathbb{E}[X_{t_n},\cov[X_{t_n}])|$  where $t_n = \sum_{k=1:n-1}h^k$ for $n=0:N$ is not our concern.
        Instead, we ultimately care about the quality of the importance sampling estimate of $P_f$ using the ensemble $\bm{x}^N$.
        To this end, it suffices that the discretization converges to the same equilibrium as the continuous process.
        Therefore rejecting the current step if the error is too large is not necessary if the discretization still moves into the right direction (the equilibrium).
        If we are sufficiently close to an equilibrium, we can apply Theorem \ref{thm:approx_non_gaussian_target} and Remark \ref{rem:other_theoretical_properties}.
        Indeed, we see from \eqref{eq:disc_cbs_sde_conv_to_equlibirum} that any stepsize $h>0$ gets us closer to the local equilibrium.
        Thus, we do not have to recompute the current step and carry on with the next step.
        If we are not close enough to an equilibrium we cannot apply the local convergence results of Theorem \ref{thm:approx_non_gaussian_target}.
		In this case, our best hope is to continue with the iteration to follow the dynamics of the   SDE \eqref{eq:def_cbs_sde} until we are in the vicinity of the correct attractor. 
        Again the local error is of secondary importance. 
       Thus, instead of recomputing a step in which we are not interested, we limit the increase of $s$ in dependence on $h_{\textup{opt}}^n$.
        As described in Section \ref{sec:choosing_smooothing_param}, we set $|s^{n+1} - s^n| \leq \textup{Lip}(s) h_{\textup{opt}}^n$ for the following step and carry on.
        This ensures that for large local errors the  target distribution $\hat{\tau}$ does not change significantly, which could increase the distance between the currents iteration's distribution and the attractor even more.
        In both cases it is still sensible to decrease the stepsize if the local error of the last step is large, as it ensures that the discrete dynamics follow their continuous conterpart more closesly.
        Therefore we use the optimal stepsize $h_{\textup{opt}}^n$ of the current step as the stepsize for the next step.
		For $n \in \mathbb{N}^+$ where $n$ is even we set:
		\begin{equation}
			\label{eq:stepsize_selection}
			\begin{split}
				h^{n+1}  &= \left ( \frac{1}{\textup{err}^n}\right )^{1/2}h^n.
			\end{split}
		\end{equation}

		\paragraph{Starting Stepsize}
		The parameter $\beta$ does not need to be initialized and the initial value of $s$ is determined by a boundary condition. 
		However, it is not obvious how to appropriately choose the stepsize $h^1>0$.
		Fortunately, there are established routines to avoid the choice of very bad values for $h^1$, cf. \cite[Ch. II.4]{ErnstHairer1993}.
		We also employ such a routine in our algorithm because the overhead of one extra ensemble update \eqref{eq:cbs_ensemble_update} is cheaper than the consequence of a very badly chosen initial stepsize.
		Let us sketch the method described in  \cite[Ch. II.4]{ErnstHairer1993}.
		The idea is to make a small Euler step and approximate the derivative of the right hand side of \eqref{eq:ode_moments_of_sde}, which we will denote by $g:\mathbb{R}^m \rightarrow \mathbb{R}^m$.
		The reason for approximating $g^\prime(\theta^0)$ is that from \cite[Ch. II. 2]{ErnstHairer1993} we know that there is a $K >0$ such that
		\begin{equation*}
			\left \| \colvec{2}{\mathbb{E}[X^1]}{\mattovec(\cov[X^1])} - \colvec{2}{\mathbb{E}[X_{h^1}]}{\mattovec(\cov[X_{h^1}])} \right \|  \approx K (h^{1})^2 g^\prime(\theta^0)
		\end{equation*}
		for any norm $||\cdot||$ on $ \mathbb{R}^m$.
		We use a norm natural in the context of error control; namely the norm $|\cdot |_\Gamma$ with 
		\begin{equation*}
			\Gamma_{i,j} = 
			\begin{cases}
				m\left( \epsilon_{\textup{Target}} + \epsilon_{\textup{Target}} |\theta^0_j|  \right), & i = j,         \\
				0,                                                                                     & i\neq j, \\
			\end{cases}
			\textup{ for } i,j=1:m.
		\end{equation*}
		Now we come to the initial Euler step. That is, we compute $\theta^1$ using the stepsize 
		\begin{equation*}
			h^1_0 = \frac{1}{100} \frac{|\theta^0|_\Gamma}{| g(\theta^0) |_\Gamma}.
		\end{equation*}
		This step is small in the sense that the size of the explicit Euler increment $h^1_0 g(\theta^0)$ is only a fraction of the size of the initial value $\theta^0$.
		To estimate the derivative of $g$, we employ a forward difference scheme and thus need to evaluate $g$ in $\theta^1$ (this is the overhead of the starting stepsize selection).
		Our second guess for $h^1$  is based on the theoretical form of the local error. We compute $h^1_1$ by solving the equation
		\begin{equation*}
			(h^1_1)^2 \max\left( \frac{|g(\theta^1) - g(\theta^0)|_\Gamma}{h^1_0},| g(\theta^0) |_\Gamma \right) = \frac{1}{100}.
		\end{equation*}
		Finally, the authors of \cite{ErnstHairer1993} propose to use the initial stepsize 
		\begin{equation}
			\label{eq:starting_stepsize}
			h^1 = \max(100 h^1_0, h^1_1).
		\end{equation}
					
		\subsection{A stopping criterion}
		\label{sec:stopping_criterion}
		In this section, we describe a criterion to stop  the time-discrete particle approximation \eqref{eq:cbs_ensemble_update}.
		The criterion is based on two checks, a convergence and a divergence check.
		The first stops the algorithm because we deem the current estimate to satisfy a specified tolerance. 
		The second check on the other hand stops the algorithm because we deem it implausible to reach the specified tolerance at all.
		\paragraph{The Convergence Check}
		This check is also used for sequential importance sampling \cite{Papaioannou2016} and the EnKF for rare event estimation \cite{Wagner2022}.
		Namely, we stop if the empirical coefficient of variation of the importance sampling weights $\bm{r}^N$ is at most  $\Delta_{\textup{Target}}$ (the user specified parameter from Section \ref{sec:choosing_smooothing_param}).
		The weights $\bm{r}^N$ are defined by
		\begin{equation}
			\label{eq:importance_weights}
			r^N_j := \frac{\pi(x^N_j) \mathds{1}_{\{G \leq 0\} } (x^N_j)}{ \mu^N(x^N_j)}, \quad j = 1:J.
		\end{equation}
		Here $\mu^N$ is the Gaussian density fitted to the sample $\bm{x}^N$, cf. Section \ref{sec:underlying_idea}, which we use for the importance sampling estimate 
		\begin{equation}
			\label{eq:cbree_is_esimate}
			\widehat{P}_f^N = \frac{1}{J}\sum_{j=1:J}r^N_j.
		\end{equation}
		This stopping criterion is justified by the following consideration.
		For an independent sample $\bm{x}^N$ we would have:
		\begin{equation*}
			\left\| {(P_f - \widehat{P}_f^N) }/{P_f}\right\|_{L^2} = \variation(\widehat{P}_f^N ) \approx \frac{\widehat{\variation}(\bm{r}^N)}{\sqrt{J}} \leq \frac{\Delta_{\textup{Target}}}{\sqrt{J}}.	
		\end{equation*}
		Although our sample is not independent, we can use the same criterion. 
		Concretely, we say that the convergence check is passed if
		\begin{equation}
			\label{eq:convergence_check}
			\widehat{\variation}(\bm{r}^N) \leq \Delta_{\textup{Target}}.
		\end{equation}
		\paragraph{The Divergence Check}
		This secondary check covers the possibility that the convergence check in \eqref{eq:convergence_check} is never triggered because the value $\Delta_{\textup{Target}}$ provided by the user is chosen too small.
		This can easily be the case for a problem where the Laplace approximation of the optimal density $\mu_{\textup{opt}}$ is a poor choice for importance sampling. As we have pointed out
		in Section \ref{sec:underlying_idea} in this case our algorithm might struggle to produce a sample suited for importance sampling.
		But also in this instance, we would like to provide some estimate of  ${P_f}$.
		In our experience, the relative error $\displaystyle{|P_f - \widehat{P_f}^n|}/{P_f}$, as well as its proxy $\widehat{\variation}(\bm{r}^n)$,  tends to show the following qualitative behavior.
		The error decreases more or less immediately with the first iteration. After some time, the error increases again. It oscillates for some iterations at a higher level before reducing again to a level comparable to the accuracy of the best estimates seen so far during the simulation.
		This wave pattern is then repeating itself.
		If we assume that the best estimates of each wave are more or less of the same quality, it makes sense to stop the simulation if the error estimate increases for the first time and has not been smaller than $\Delta_{\textup{Target}}$. 
		
		For this reason, we introduce the parameter $N_{\textup{obs}} \geq 2$.
		This defines an \emph{observation window}, i.e., we will have closer look at the $N_{\textup{obs}}$ last ensembles.
		If $n \geq N_{\textup{obs}} $, we perform the divergence check.
		We fit a linear function to the points $(k, \widehat{\variation}(\bm{r}^k))$, ${k=n-N_{\textup{obs}}+1:n}$, by a least square approximation and stop the simulation if the slope of this line is positive.
		If we stop at step $N$ because of the divergence check, it would not be surprising if $\widehat{P}_f^N$ is not the best estimate.
		Instead, given the last $N_{\textup{obs}}$ importance sampling estimates, we want to compute an optimal estimate.
		This problem is investigated in \cite{Owen2013}.
		The author advises computing the final estimate as a convex combination of the last $N_{\textup{obs}}$ estimates.
		\begin{equation}
			\label{eq:lin_comb_estimates}
			\widehat{P}_f = \sum_{n=N-N_{\textup{obs}}+1:N} \omega_n \widehat{P_f^n}.
		\end{equation}
		In our case, we do not expect the quality of the last $N_{\textup{obs}}$ estimates to improve as $n$ increases because the divergence check has just been triggered.
		For this reason, we follow the recommendation of  \cite{Owen2013} and use uniform weights, i.e. $\omega_n = N_{\textup{obs}}^{-1}$ for $n = N-N_{\textup{obs}}+1:N$.

		\section{EnKF vs. CBREE}
		\label{sec:enkf_vs_cbree}
		In this section we compare the CBREE method, Algorithm \ref{sec:algorithm}, to the EnKF for rare event estimation in \cite{Wagner2022}.
		Both methods have historically first been used for Bayesian inverse problems.
		Indeed, the paper \cite{Carrillo2022}, which introduced the consensus\--based sampling dynamics \eqref{eq:def_cbs_sde}, applied them to Bayesian inversion, and the EnKF method in \cite{Wagner2022} is explicitly derived from an existing method for inverse problems based on the well known Kalman filter, cf. \cite{Iglesias2013, Schillings2017}.
		This comparison is of interest as both methods use the idea of moving an initial sample along the trajectory of an SDE to obtain a new sample for use with importance sampling.
		
		Firstly, we compare the stochastic processes that govern the respective sample updates.
		Let us state the counterpart of \eqref{eq:def_cbs_sde} in the EnKF method.
		If we start with the EnKF particle update \cite[(3.5)]{Wagner2022}, we  can use the results from \cite{Ernst2015, Schillings2017} to  take the mean field limit and obtain the time discrete stochastic process 
		\begin{equation*}
			\begin{split}
				X^0 &\sim \mathcal{N}(0,I_d), \\
				X^{n+1} &= X^n - \cov[X^n, G^n] \cov[G^n, G^n]^{-1}G^n, \\
				G^n &= \max(G(X^n), 0)+\xi^n, \\
				(\xi^n)_{n \in \mathbb{N}} & \stackrel{\textup{i.i.d.}}{\sim}\mathcal{N}(0,h^{-1}),
			\end{split}
		\end{equation*}
		where $\cov[X,Y]=\mathbb{E}[(X-\mathbb{E}[X]) (Y-\mathbb{E}[Y])^T]$ for random vectors $X$ and $Y$ with finite second moments.
		Then we take the limit $h\rightarrow 0$ as described in \cite[Sec.~3]{Schillings2017} to obtain the following McKean--Vlasov SDE,
		\begin{equation}
			\label{eq:enkf_sde}
			\begin{split}
				X_0 &\sim \mathcal{N}(0,I_d), \\
				dX_t &= - \cov[X_t, \max(G(X_t),0)](\max(G(X_t),0)dt + dW_t).
			\end{split}	
		\end{equation}
		Note that the diffusion term above consists of a $d$-dimensional vector multiplied by a scalar Wiener process whereas the diffusion of the consensus--based sampling dynamics in \eqref{eq:def_cbs_sde} is given by a matrix times a $d$-dimensional Wiener process.
		To get an intuition behind these dynamics we consider the following approximation of \eqref{eq:enkf_sde}:
		\begin{equation}
			\label{eq:approx_enkf_sde}
			d \widetilde{X}_t =  - \cov[\widetilde{X}_t,\widetilde{X}_t ]\nabla \max(G(\widetilde{X}_t),0)(\max(G(\widetilde{X}_t),0)dt + dW_t).
		\end{equation}
		Here $\nabla$ should be thought of as a weak derivative.
		The process $\widetilde{X}_t$ approximates the process $X_t$ from \eqref{eq:enkf_sde} and they are identical (in distribution) if  the map $x \mapsto \max(G(x),0)$ is linear.
		To see this, note that $$\max(G(x),0) = [\nabla \max(G(x),0)]^Tx$$ implies
		    $\cov[X, \max(G(X),0)] = \cov[X, X]\nabla \max(G(X),0)$
		for any random vector $X$ whose second moments exist.
		The following interpretations of  \eqref{eq:approx_enkf_sde} also hold for the original process in \eqref{eq:enkf_sde} if only $G$ itself is linear, cf. \cite[Sec. 4]{Wagner2022}.
		In \eqref{eq:approx_enkf_sde} we recognize the gradient descent of the functional ${l(x) = \frac{1}{2} (\max(G(x),0))^2}$ preconditioned by $ \cov[\widetilde{X}_t,\widetilde{X}_t]$ and with a stochastic perturbation added to the data misfit $\max(G(\widetilde{X}_t),0)$ in form of the scalar Wiener process $W_t$.
		That is, the process $\widetilde{X}_t$ defined by the right hand side of \eqref{eq:approx_enkf_sde} drifts towards the failure domain $\{G \leq 0\}$ if it is outside the failure domain, while there is no drift in the failure domain itself.
		Indeed, the same result is also proven for the process \eqref{eq:enkf_sde} in \cite{Wagner2022} for the special case of no perturbation and a linear limit state function $G$.
		
\begin{figure}
    \begin{center}
        \includegraphics[width=.33\textheight]{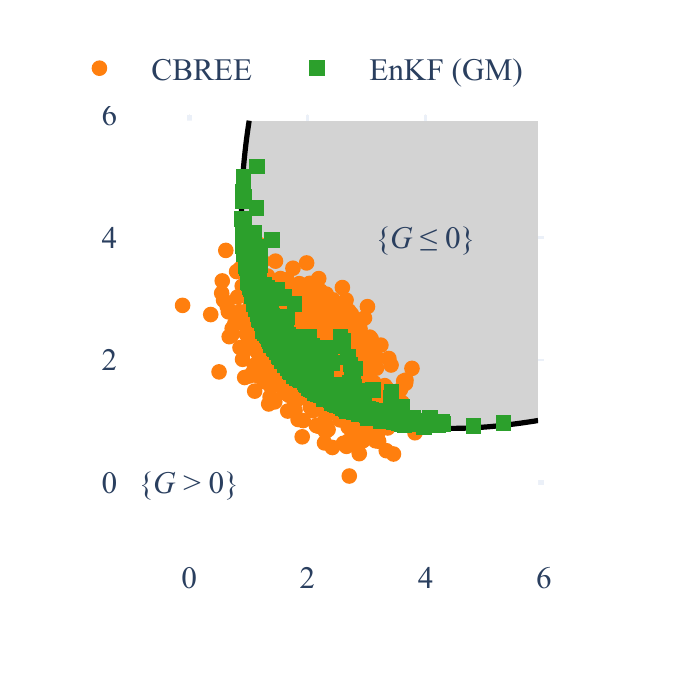}
    \end{center}
    \caption{The final ensembles of the EnKF and CBREE methods applied to the convex limit state function $G(x)= \frac{(x_1-x_2)^2}{10} - \frac{x_1 +x_2}{\sqrt{2}} + \frac{2}{5}$.  Note that for the EnKF method we show the last particle ensemble of the internal iteration not the sample used for importance sampling. Each method used $J=1000$ samples and the stopping criterion $\Delta_{\text{Target}}$ = 1. The CBREE method performed no divergence check, used the stepsize control $\epsilon_{\text{Target}}=0.5$ and controlled the increase of $s$ with $\text{Lip}(s) = 1$.}
    \label{fig:scatter_plot}
\end{figure}

		The intuitions we have developed for the EnKF dynamics are inherited by the time-discrete EnKF method itself.
		Analogously to the CBREE method, the EnKF method transforms an initial sample $\bm{x}^0 \sim \mathcal{N}(0,I_d)$ in $N$ steps into the final sample $\bm{x}^N$, which is used as the basis of an importance sampling estimate of \eqref{eq:def_pf}.
		However, the respective final samples of both methods can differ significantly due to the different underlying dynamics.
		In the EnKF approach the final sample hugs the failure surface $\{G = 0\}$ as the points of the initial sample $\bm{x}^0$ approximately follow the flow of the gradient of $l(x) = \frac{1}{2} (\max(G(x),0))^2$ up to the failure surface.
		On the other hand, the final sample of the CBREE method is Gaussian centered close to the global maximum of the function $x \mapsto I(G(x), s^N)\pi(x)$ which approximates $\mu_{\textup{opt}}$  and is therefore much less flexible in its spatial distribution.
        This phenomenon is visualized in Figure \ref{fig:scatter_plot}.
        
	Now we come to our second point.
    Unfortunately, the distribution of the final EnKF sample is not known in closed form.
    Hence we cannot use it directly for importance sampling.
		Instead, one has to fit a distribution, say a Gaussian mixture, to the final sample.
		Then one can resample once from that fitted distribution to obtain a sample for importance sampling, cf. \cite{Wagner2022}.
		In contrast, the distribution of the final sample associated with the CBREE approach is known: it is Gaussian provided that the initial sample is Gaussian, cf. \cite{Carrillo2022}.

%% file: numerical_experiments.tex
\section{Numerical Experiments}
\label{sec:numerical_experiments}
In this section we present several numerical experiments.
We study how the parameter choices influence the performance of the CBREE method.
Furthermore, we compare our algorithm to the SIS method in \cite{Papaioannou2016} and the EnKF method in \cite{Wagner2022}.
These benchmark methods come with different options. Here we will employ the following versions:
\begin{itemize}
	\item \emph{EnKF (GM)} denotes the EnKF method \cite{Wagner2022} using a single Gaussian for the importance sampling step. \enquote{GM} is the abbreviation of \emph{Gaussian mixture}. 
        As we have pointed out in the introduction of Section \ref{sec:algorithm}, we only consider unimodal problems; hence the Gaussian mixture has only one component.
    \item \emph{EnKF (vMFNM)} denotes the EnKF method \cite{Wagner2022} using a \emph{von\--Mises\--Fisher\--Nakagami} (vMFN) distribution for the importance sampling step. 
	The vMFN distribution will be described in more detail in Section \ref{sec:performance_in_higher_dimensions}.
	Again, EnKF (vMFNM) uses only one vMFN component instead of a mixture of vMFN distributions as in \cite{Wagner2022}. 
\item\emph{SIS (GM)} denotes the SIS method using a normal distribution as the proposal density for its \emph{Markov chain Monte Carlo} subroutine.
	\item\emph{SIS (vMFNM)} denotes the SIS method using a vMFN distribution as the proposal density in the MCMC step.
\end{itemize}
The EnKF and SIS methods use the same stopping criterion as the CBREE method in equation \eqref{eq:convergence_check}. For all four benchmark methods, we fix the value $\Delta_{\text{Target}}=1$.

In the following, we present three experiments.
First, we consider two low-dimensional rare event estimation problems to benchmark our method and study the behavior of the stopping criterion presented in Section \ref{sec:stopping_criterion}.
Then, we study the performance of our method in higher dimensions.
As observed in previous studies \cite{Papaioannou2016, Wagner2022}, we expect that the SIS and EnKF methods using the vMFN distribution perform better in higher dimensions than their counterparts using Gaussian mixture (GM) distributions.

We measure the performance of a rare event estimation algorithm using the \emph{relative efficiency} discussed in \cite{chan2023bayesian}. 
To motivate the idea recall that each method outputs an estimate $\widehat{P}_f$ which is a random variable.
Running the method $K$ times produces the independent realizations $\widehat{P}_f^1, \ldots, \widehat{P}_f^K$.
Based on those estimates we compute the empirical mean squared error and average cost associated with the random variable $\widehat{P}_f$:
\begin{equation*}
    \MSE(\widehat{P}_f) 
    =
    \frac{1}{K} \sum_{k=1:K} (\widehat{P}_f^k - P_f)^2,\quad  \cost(\widehat{P}_f) 
    =
    \frac{1}{K} \sum_{k=1:K} \cost(\widehat{P}_f^k).
\end{equation*}
As it is common in the context of rare event estimation, we define $\cost(\widehat{P}_f^k)$ as the number of limit state function evaluations during the computation of the estimate $\widehat{P}_f^k$.
The relative efficiency is in turn defined as 
\begin{equation}
    \relEff(\widehat{P}_f) 
    = 
    \frac{P_f(1-P_f)}{\MSE(\widehat{P}_f) \times \cost(\widehat{P}_f)}.
    \label{eq:rel_eff}
\end{equation}
This quantity can be readily interpreted.
If we define the efficiency of an algorithm producing $\widehat{P}_f$ as $1/(\MSE(\widehat{P}_f)\times \cost(\widehat{P}_f))$ \cite{l1994efficiency}, i.e., the smaller the error and (or) the cost of the final estimate the more efficient is the underlying algorithm, then we can analytically compute the efficiency of Monte Carlo sampling.
The latter turns out to be $1/(P_f(1-P_f))$.
Thus, the relative efficiency measures how many times more efficient an algorithm is compared to Monte Carlo sampling.

\subsection{Nonlinear Oscillator}
\label{sec:nonlinear_oscillator}
This problem is based on the model of a nonlinear oscillator and is taken from \cite[Ex. 4.3]{Cheng2023}.
The limit state function is defined in terms of the variables $x = [M, c_1, c_2, r , F_1, t_1]^T$ and reads
\begin{equation}
    \label{eq:nonlinear_oscillator}
    G: \mathbb{R}^6 \rightarrow \mathbb{R},\quad x \mapsto 3r 
    - \left| 
    \frac{2F_1}{M \omega_0^2}
    \sin \left(
        \frac{t_1 \omega_0}{2}
    \right)
        \right|, \quad \omega = \sqrt{\frac{c_1 + c_2}{M}}.
\end{equation}
The corresponding probability distribution $\pi$ is an uncorrelated normal distribution $\mathcal{N}(a, A)$ defined by the parameters
\begin{equation*}
    a = [1, 1, 0.1, 0.5, 0.3, 1]^T, \quad A = \diag(0.05^2, 0.1^2, 0.01^2, 0.05^2, 0.2^2, 0.2^2).
\end{equation*}
We use the reference value $P_f = 6.43 \cdot 10^{-6}$, which is computed in  \cite[Ex. 4.3]{Cheng2023} with a Monte Carlo simulation using $10^9$ samples.

\begin{figure}
    \begin{center}
        \includegraphics[width=\textwidth]{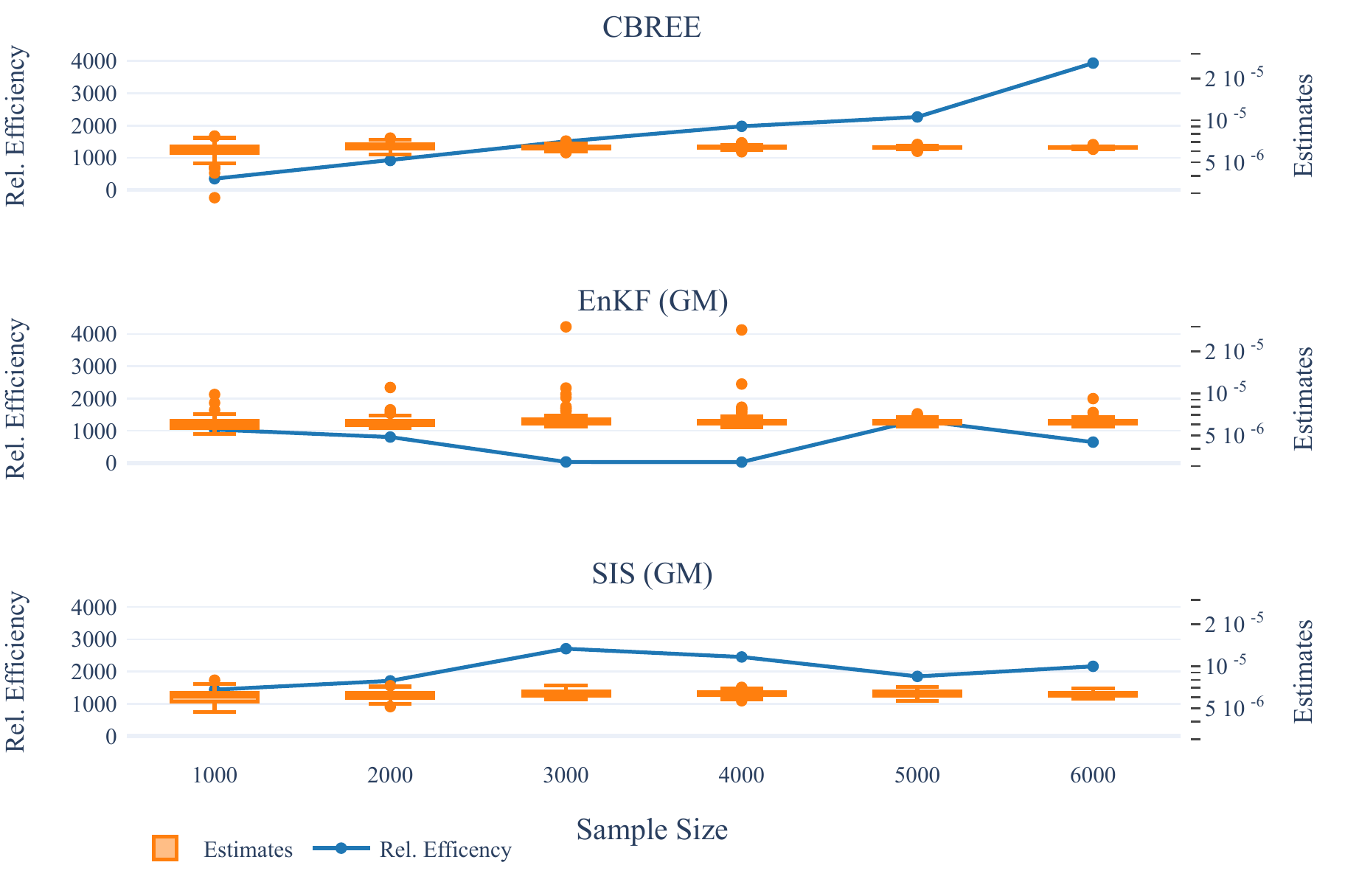}
    \end{center}
    \caption{\protect\input{eficiency_plot_nonlinear_oscillator_problem_solver_vs_sample_size_desc.tex}}
    \label{fig:benchmark_oscillator}
\end{figure}

Now we compare the CBREE method with the benchmark methods.
The result is given in Figure \ref{fig:benchmark_oscillator}. 
We highlight two observations from this figure.
Firstly, we see that the relative efficiency is very sensitive to outlier estimates of $P_f$.
This can be seen in Figure \ref{fig:benchmark_oscillator} for the CBREE method and sample size $1000$ as well as for the EnKF (GM) method and the samples sizes $2000$ to $4000$.
This is not surprising as this measure depends on the mean of the squared errors which itself is known to have the same sensitivity.
Furthermore, the EnKF method displays the largest outliers.
The percentage of outliers observed for the EnKF method has been reported in \cite{Wagner2022} for various problem settings.
Our second observation is that the CBREE method tends to be more efficient for larger sample sizes $J$.
Indeed in this case our method is even more efficient than the benchmark methods for $J \in \{5000, 6000\}$.

\begin{figure}
    \begin{center}
        \includegraphics[width=\textwidth]{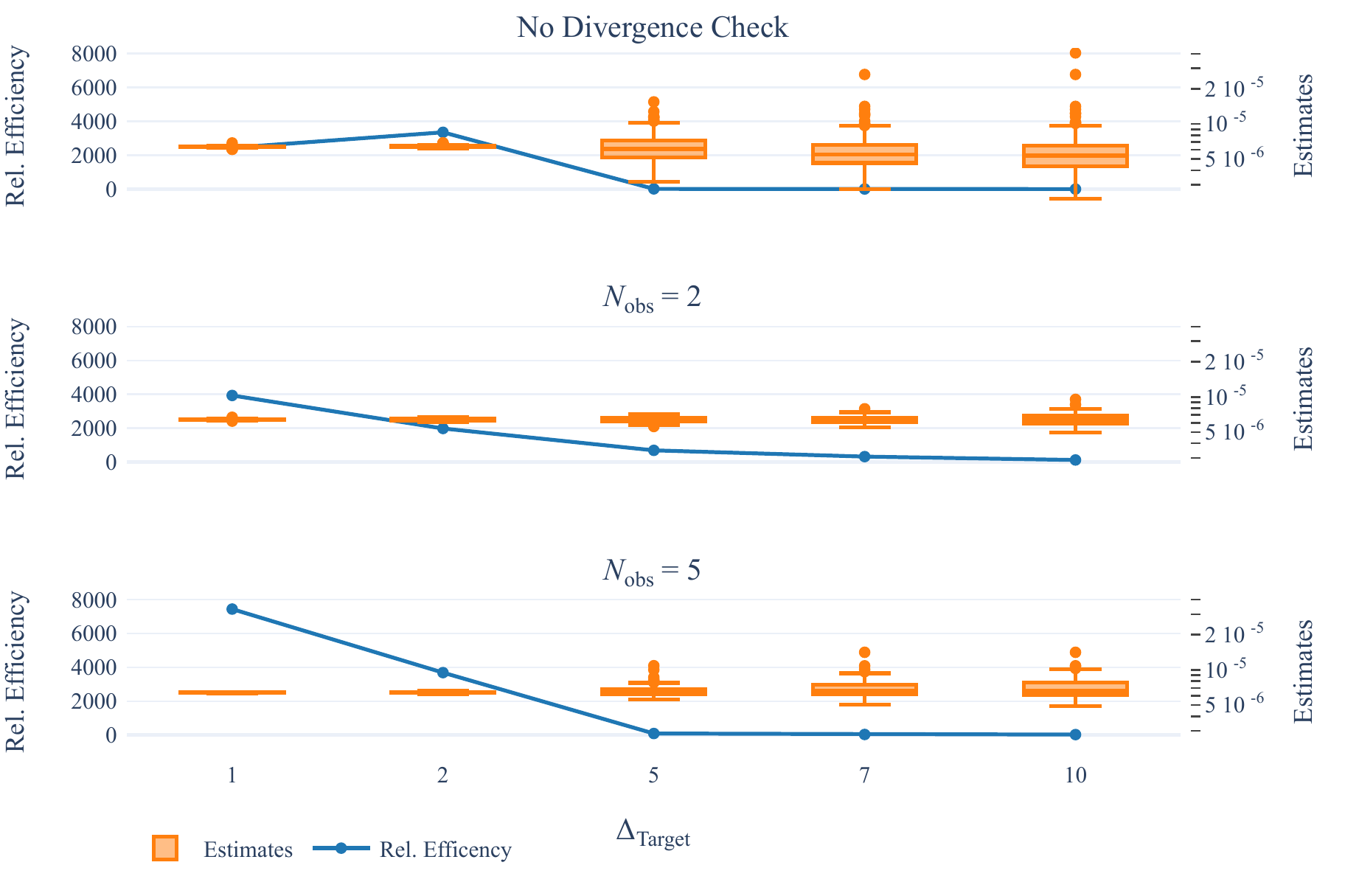}
    \end{center}
    \caption{\protect\input{eficiency_plot_nonlinear_oscillator_problem_delta_tgt_vs_obs_window_desc.tex}}
    \label{fig:parameter_study}
\end{figure}

We also present a parameter study for the stopping criterion, cf. Section \ref{sec:stopping_criterion}, based on the Nonlinear Oscillator Problem.
For this we fix the sample size $J=6000$ and vary the parameters $\Delta_{\textup{Target}}>0$ and $N_{\textup{obs}}\geq 0$.
Here the case $N_{\textup{obs}} = 0$ corresponds to skipping the divergence check in line \ref{alg_line:div_check} of Algorithm \ref{alg:cbree}.
The results are depicted in Figure \ref{fig:parameter_study}.
The figure shows that the efficiency of the CBREE methods decreases if the parameter $\Delta_{\text{Target}}$ increases.
This is a general trend which is independent of the divergence check and the parameter choice of  $N_{\textup{obs}}$.
From the boxplots of the probability estimates we can discern that this trend correlates with a widening of the spread of the estimate produced by the CBREE method.
Furthermore, we see that foregoing the divergence check tends to increase the variance of the final error estimate $\widehat{P}_f$ compared to the cases where $N_{\textup{obs}} \geq 2 $.
In the latter case the occurance of outlier estimates appears to be less likely, cf. the boxplots in Figure \ref{fig:parameter_study} for $\Delta_{\textup{Target}} \geq 5$.
But whereas the relationship between $\Delta_{\textup{Target}}$ and the efficiency of the method appears to be monotone, the same cannot be said
about the effect of the parameter $N_{\textup{obs}}$ as we see by comparing the choices  $N_{\textup{obs}} = 2 $ and $N_{\textup{obs}} = 5$. 
We can see that choosing the observation window too large can decrease the efficiency.
In our experience, the choice $N_{\textup{obs}} = 2 $ yields the best results.

\subsection{Flowrate Problem}
\label{sec:flowrate_problem}
The evaluation of this problem's limit state function involves the approximate solution of a partial differential equation (PDE).
We have taken this example from \cite[Sec. 5.3]{Wagner2021}.
Let $(\Omega, \mathcal{F}, \mathbb{P}) $ be  a probability space and
consider the  1D boundary value problem with a random diffusion coefficient
\begin{equation}
	\label{eq:duffsuion_pde}
	\begin{split}
		&\frac{d}{dy}\left (-a(y,\omega) \frac{d}{dy} u(y, \omega) \right ) = 0 \quad \forall y \in (0,1), \\
		& u(0, \omega) = 1,\, u(1, \omega)=0 \quad \text{for $\mathbb{P}$-a.e. } \omega \in \Omega.
	\end{split}
\end{equation}
Here the diffusion coefficient $a:(0,1) \times \Omega \rightarrow \mathbb{R}$ is a log-normal random field.
This means that for each finite set $(y_i)_{i=1:d} \subset (0,1)$ the random vector $[\log a(y_1), \ldots, \log a(y_d)]^T$ follows the multivariate normal distribution with first two moments depending on $(y_i)_{i=1:d}$, cf. \cite[Ch. 7.1]{Lord2014}.
The distribution of $a$ is completely determined by the mean function $y \mapsto \mathbb{E}[\log a(y, \cdot)]$ and the covariance function $(y_1,y_2) \mapsto \cov[\log a(y_1, \cdot), \log a(y_2, \cdot)]$. 
For our example, we set
\begin{equation*}
	\begin{split}
		\mathbb{E}[\log a(y, \cdot)] & = 0.1 \quad \forall y \in (0,1), \\
		\cov[\log a(y_1, \cdot), \log a(y_2, \cdot)] &= 0.04 e^{-10/3 |y_1-y_2|} \quad \forall y_1,y_2 \in (0,1). 
	\end{split}
\end{equation*}
One can show that the stochastic PDE in \eqref{eq:duffsuion_pde} has a unique weak solution $u$, \cite[Thm. 9,9]{Lord2014}.
In order to approximately sample from the solution of \eqref{eq:duffsuion_pde}, one has first to sample from an approximation of the random field $a$ and then approximate the solution of the PDE, which is deterministic once the diffusion coefficient has been sampled.
The first part is commonly done by truncating a  Karhunen–Lo\`{e}ve expansion of $\log a$, cf. \cite[Ch. 7.4]{Lord2014}.
Let  $d \in \mathbb{N}^+$ and think of the random variables $(\xi_i)_{i=1:d} \stackrel{\text{i.i.d.}}{\sim} \mathcal{N}(0,1)$ as the  components of the random vector $\xi\sim \mathcal{N}(0, I_{d})$.
It is possible to analytically calculate values $(\lambda_i)_{i=0:d} \subset \mathbb{R}$ and functions $(v_i)_{i=1:d} \subset \mathcal{C}(D, \mathbb{R})$ such that we have
\begin{equation*}
	\log a_{d}(y, \xi(\omega)) :=  \lambda_0 + \sum_{i=1:d} \lambda_i v_i(y) \xi_i(\omega) \rightarrow \log a(y, \omega)  \text{ in } L^2(\Omega, L^2(0,1)),\quad d \rightarrow \infty,
\end{equation*}
cf. \cite[Ex. 7.55 ]{Lord2014}.
We set $d=10$ and replace the diffusion coefficient $a$ in \eqref{eq:duffsuion_pde} by the approximation $a_d$.
For a given realization $x = \xi(\omega)$ the stochastic PDE in \eqref{eq:duffsuion_pde} turns into a deterministic boundary value problem.
We approximate the solution of this problem using piecewise linear continuous finite elements on a uniform grid on $(0,1)$ with mesh size $h = 2^{-6}$ and denote the approximation by $ u_h(y, x)$.
For the theory of the finite element method see e.g. \cite[Ch. 8]{Hackbusch2017}.
We arrive at the limit state function
\begin{equation}
	\label{eq:diffusion_lsf}
    G: \mathbb{R}^{10} \rightarrow \mathbb{R},\quad x \mapsto 1.7 + a_d(1,x) \frac{d}{dy} u_h(1, x)
\end{equation}
That means $\{G \leq 0\}$ is the event of the flowrate $-a_d(1,x) \frac{d}{dy} u_h(1, x)
$ exceeding the threshold $1.7$ at $y =1 $.
The reference value used in the following is $P_f=3.026 \cdot 10^{-4}$.
We computed this value using a Monte Carlo simulation with $10^7$ samples.

\begin{figure}
    \begin{center}
        \includegraphics[width=\textwidth]{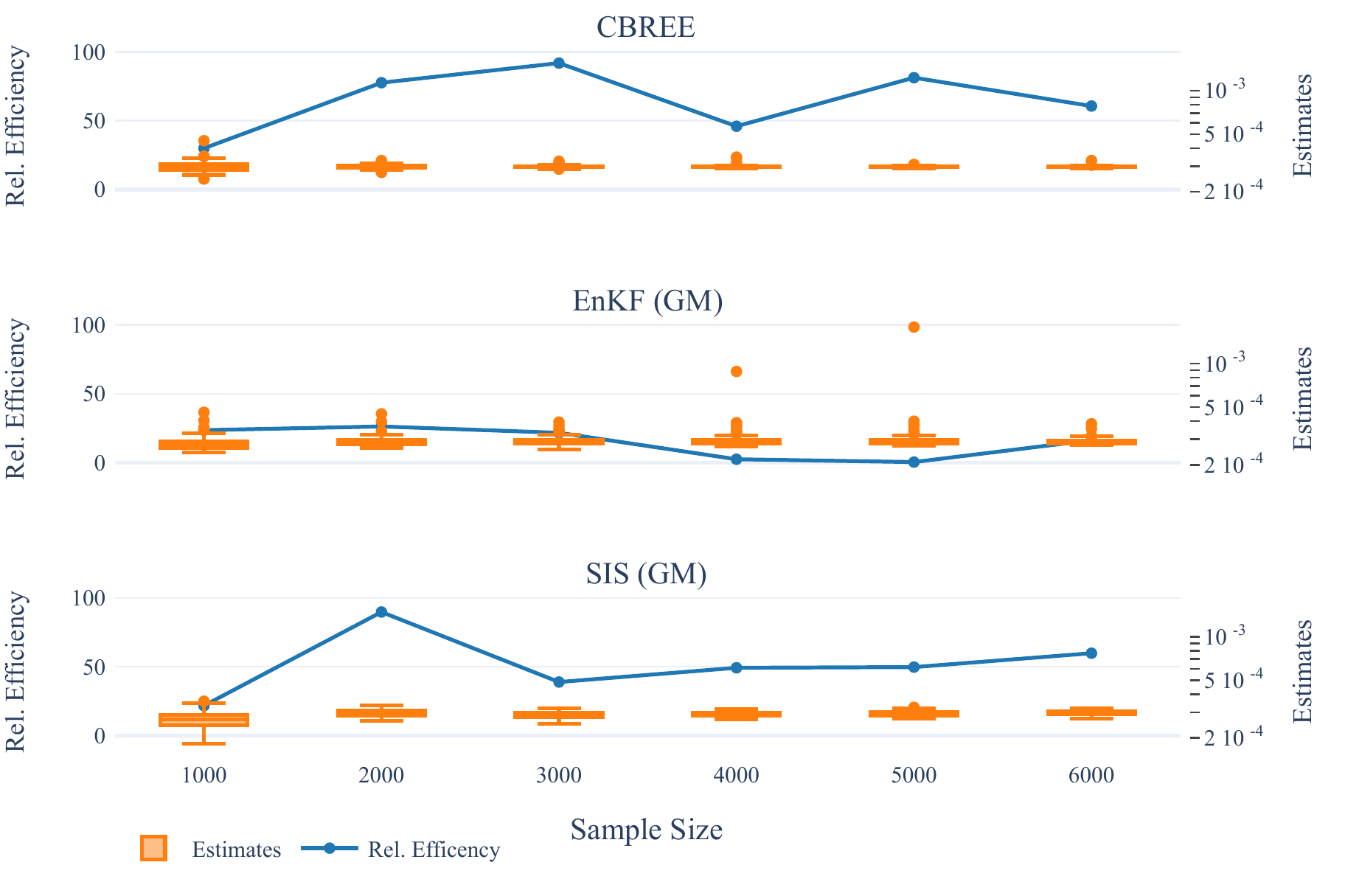}
    \end{center}
    \caption{\protect\input{eficiency_plot_flow-rate_problem_solver_vs_sample_size_desc.tex}}
    \label{fig:flowrate_benchmark}
\end{figure}

The results of this experiment are given in Figure \ref{fig:flowrate_benchmark}.
Again we observe the sensitivity of the relative efficiency to outliers.
This can be seen for instance for the EnKF (GM) method using the sample sizes $4000$ and $5000$.
Furthermore, we note that in this case, the CBREE method often outperforms the benchmark methods.

\subsection{Performance in Higher Dimensions}
\label{sec:performance_in_higher_dimensions}
We now study the performance of our method for a higher dimensional problem.
We choose the Linear Problem whose failure domain is bounded by a hypersurface in $\mathbb{R}^d$.
Due to the simple form of the failure domain, we can compute $P_f$ explicitly for any dimension $d$.
The limit state function reads
\begin{equation}
	\label{eq:linear_lsf}
	G: \mathbb{R}^d \rightarrow \mathbb{R}, \quad
	x \mapsto c -  \frac{1}{\sqrt{d}} \sum_{i=1:d} x_i.
\end{equation}
One can check that $P_f = \Phi(-c)$ where $\Phi$ is the cumulative distribution function of the standard normal distribution $\mathcal{N}(0,1)$.
In the following  we use $c = 3.5$ such that $P_f = 2.32 \cdot 10^{-4}$ for any $d$.
Here we solve the problem for $d \in \{2,50\}$ with our method and the benchmark methods. 

As we have already mentioned at the beginning of Section \ref{sec:numerical_experiments}, both benchmark methods come in two different versions.
The versions depicted so far, SIS (GM) and EnKF (GM), are based on sampling from a Gaussian.
The sampling from a Gaussian in higher dimensions leads to problems in the context of importance sampling, cf. \cite{Katafygiotis2008} for a geometrical explanation of this phenomenon.
The von\--Mises\--Fischer\--Nakagami distribution (vMFN) on the other hand can be thought of as a Gaussian $\mathcal{N}(a, A)$, where $A$ is proportional to the identity $I_d$, but whose radial component has elongated tails. 
This change precisely addresses the issues presented in \cite{Katafygiotis2008}.
For more information on this distribution and how it can be fitted to a given sample, see \cite{Papaioannou2019}.
The replacement of the Gaussian by a von\--Mises\--Fischer\--Nakagami distribution for importance sampling leads to the methods SIS (vMFNM) and EnKF (vMFNM) which perform in general better in higher dimensions \cite{Papaioannou2016, Wagner2022}.

Interestingly, we can use the vMFN distribution for the CBREE method.
We replace every ensemble $\bm{x}^0, \bm{x}^1, \ldots$ by a resampled vMFN proxy before evaluating the limit state function.
Formally, let ${\texttt{vmfn\_resample}:\bm{x} \mapsto (\bm{y}, \mu)}$ a subroutine that takes in a sample $\bm{x} = (x_j)_{j=1:J} \subset \mathbb{R}^d$. This algorithm fits a vMFN density $\mu$ onto the input sample and produces a new ensemble $\bm{y} \sim \mu$.
Finally, the sample $\bm{y}$ and the density $\mu$ are returned.
Then we can add a new step between Line 2 and 3 of Algorithm \ref{alg:cbree} as described in Algorithm \ref{alg:cbree_vmfn}.
\begin{algorithm}
\caption{Consensus-based sampling for rare event estimation in high dimensions, CBREE (vMFN).}
    \label{alg:cbree_vmfn}
    \begin{algorithmic}[1]
        \STATE{Compute initial stepsize $h^1$ according to \eqref{eq:starting_stepsize}}
        \FOR{$n \in \mathbb{N}$}
        \STATE{$(\bm{x}^{n}, {\mu^n}) \gets \texttt{vmfn\_resample}(\bm{x^n})$}
        \STATE{$\widehat{P}_f^n = \frac{1}{J}\sum_{j=1:J} \frac{\pi(x^n_j) \mathds{1}_{\{G \leq 0\} } (x^n_j)}{ \mu^n(x^n_j)}$}
        \STATE{Execute lines \ref{alg_line:begin_cbree_vmfn_recycle} to \ref{alg_line:end_cbree_vmfn_recycle} from Algorithm \ref{alg:cbree}}
        \ENDFOR
    \end{algorithmic}
\end{algorithm}

\begin{figure}
    \begin{center}
        \includegraphics[width=\textwidth]{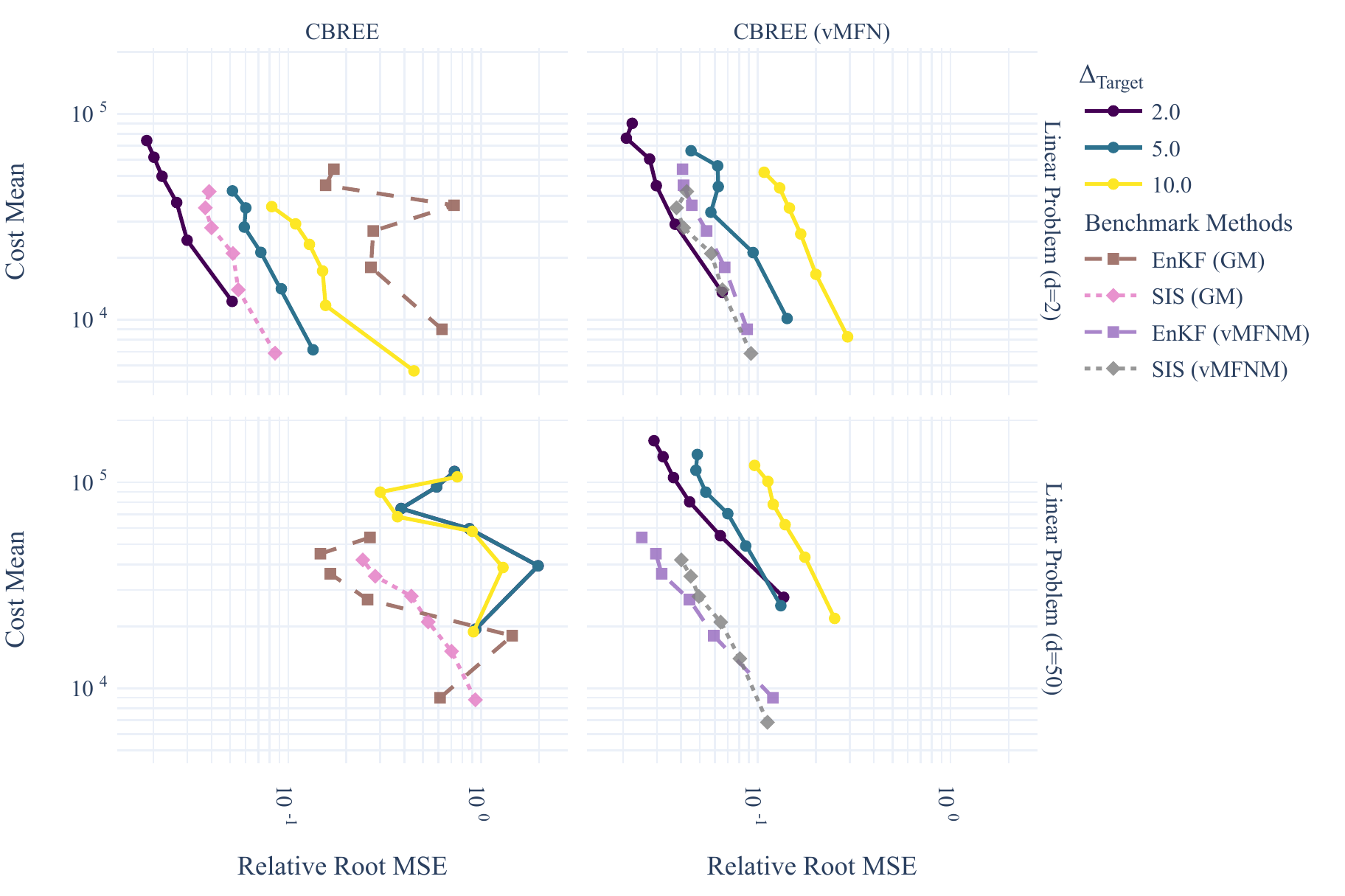}
    \end{center}
    \caption{\protect\input{linear_problems_lower_and_higher_dimensions_desc.tex}}
    \label{fig:linear_problem}
\end{figure}

We show the performance of our method and the benchmarks in Figure \ref{fig:linear_problem}.
There we plot for different sample sizes $J \in \{10^3, 2\cdot 10^3, \ldots, 6 \cdot 10^3\}$ the relative root mean squared error, $\sqrt{\MSE(\widehat{P}_f)}/P_f$, and the average cost, $\cost(\widehat{P}_f)$.
This representation highlights the convergence behavior of the methods and also informs us in which cases we can recommend the use of our method.
Having a look at the first row in Figure \ref{fig:linear_problem}, which depicts the Linear Problem for $d=2$, we see that the changes we have made in Algorithm \ref{alg:cbree_vmfn} increase the overall cost of our method.
One reason for this could be that we have changed the dynamics of the consensus\--based sampling recursion \eqref{eq:cbs_ensemble_update} which could have slowed the speed of convergence to the steady state.
Now we consider the second row where we changed the problem's dimension to $50$.
Here we can see that the original CBREE method as well as the benchmark methods based on Gaussians produce not only a large error but also struggle to converge.
The effect is most severe for the CBREE method, which did not produce a single estimate for any sample size for the parameter choice $\Delta_{\textup{Target}} = 2$ because the stopping criterion was not triggered in the first $100$ iterations.
The methods based on the von\--Mises\--Fischer\--Nakagami distribution on the other hand again display a consistent convergence behavior.

Finally, we note that the benchmark method EnKF (GM) struggles with this particular rare event estimation problem. 
As we have explained in Section \ref{sec:enkf_vs_cbree} the importance sampling estimate by the EnKF method is based on a sample obtained from the dynamics of the Kalman filter.
In particular we know that this  sample will be clustered on the hypersurface $\{G = 0\}$.
Apparently the  $d$-dimensional Gaussian fitted to this essentially $(d-1)$-dimensional sample is a poor importance sampling distribution whereas the von\--Mises\--Fischer\--Nakagami distribution performs better.

In Figure \ref{fig:linear_problem} we see that for high accuracy needs that warrant a large sample size $J$, the CBREE method outperforms the benchmarks SIS and EnKF.
This is also true in terms of the relative efficiency measure for the problems we have discussed in the previous Sections \ref{sec:nonlinear_oscillator} and \ref{sec:flowrate_problem}.
But we also see that especially for higher dimensional problems our method is more expensive than the depicted benchmark methods.

%% file: eficiency_plot_nonlinear_oscillator_problem_solver_vs_sample_size_desc.tex
Results for the Nonlinear Oscillator Problem with the CBREE method (top row), the EnKF (GM) method (middle row) and the SIS (GM) method (bottom row). 
We vary the sample size along the horizontal axis and show for each sample size two quantities: The estimate of the relative efficiency (left vertical axis) and a boxplot of the corresponding $100$ empirical estimates of the failure probability (right vertical axis). 
The other parameters of the CBREE method are $\Delta_{\text{Target}} = 1$, $N_\text{obs} = 2$ and $\epsilon_{\text{Target}} = 1$.

%% file: eficiency_plot_nonlinear_oscillator_problem_delta_tgt_vs_obs_window_desc.tex
Results for the Nonlinear Oscillator Problem with the CBREE method with $6000$ samples. We vary the parameter $\Delta_{\text{Target}}$ along the horizontal axis and the length of the observation window $N_\text{obs}$ in each row. For each parameter choice we plot an estimate of the relative efficiency (left vertical axis) and a boxplot of the corresponding $100$ empirical estimates of the failure probability (right vertical axis). The parameter $\epsilon_{\text{Target}} = 1$ is fixed.

%% file: eficiency_plot_flow-rate_problem_solver_vs_sample_size_desc.tex
Results for the Flowrate Problem ($d=10$) with the CBREE method (top row), the EnKF (GM) method (middle row) and the SIS (GM) method (bottom row). We vary the sample size along the horizontal axis and show for each sample size two quantities: The estimate of the relative efficiency (left  vertical axis) and a boxplot of the corresponding $100$ empirical estimates of the failure probability (right vertical axis). The other parameters of the CBREE method are $\Delta_{\text{Target}} = 1$, $N_\text{obs} = 2$ and $\epsilon_{\text{Target}} = 1$.

%% file: linear_problems_lower_and_higher_dimensions_desc.tex
Results for the Linear Problem with the CBREE method (first column) and the CBREE (vMFN) method (second column) using  different parameters. We vary the stopping criterion $\Delta_{\text{Target}}$ according to the colors in the legend.
The problem dimension is $d=2$ (first row) and $d=50$ (second row).
 The parameters $\epsilon_{\text{Target}} = 0.5$ and $N_\text{obs}=2$ are fixed. Furthermore we plot also the performance of the benchmark methods EnKF and SIS. Each marker represents the empirical estimates based on the successful portion of $200$ simulations.

%% file: conclusion.tex
\section{Conclusion}
\label{sec:conclusion}
In this paper we  have introduced a new algorithm for rare event estimation named CBREE.
The method is based on adaptive importance sampling together with an interacting particle system which was introduced in \cite{Carrillo2022}. 
As our new algorithm depends on several parameters we have developed strategies to choose those parameters automatically based on easy to interpret  accuracy criteria.
For this purpose we have applied well known parameter updates that are also used in other algorithms and adapted them to the situation at hand.
We have noted the structural similarities between CBREE and the EnKF method from \cite{Wagner2022}, and have compared the stochastic dynamics behind the two methods.
In numerical experiments we have seen that the performance of CBREE is comparable to other state of the art algorithms for rare event estimation, namely SIS and the EnKF.
CBREE outperforms the benchmark methods for high accuracy demands but is more expensive for higher dimensional problems. 

For future work it would be interesting to adapt the CBREE method to multimodal failure domains.
By construction it can only be applied in unimodal cases. 
However, a covariance localization ansatz as in \cite{ReichWeissmann:2021} was already applied in the EnKF method in \cite[Sec. 3.3]{Wagner2022} and could also be applied here.

\section{Data Availability}
The code and data used in this paper is available at \texttt{\url{https://github.com/AlthausKonstantin/rareeventestimation}}.